\documentclass[a4paper,UKenglish,cleveref, autoref, thm-restate]{oasics-v2021}
\title{Online Vehicle Routing with Pickups and Deliveries under Time-Dependent Travel-Time Constraints}

\titlerunning{Online VRP with pickups, deliveries and time-dependent travel-times} 

\author{Spyros {Kontogiannis}}
	{Computer Engineering and Informatics Department, University of Patras, Greece
	\and Computer Technology Institute and Press ``Diophantus'', Greece}
	{spyridon.kontogiannis@upatras.gr}
	{https://orcid.org/0000-0002-8559-6418}
	{}%(Optional) author-specific funding acknowledgements

\author{Andreas {Paraskevopoulos}}
	{Computer Technology Institute and Press ``Diophantus'', Greece}
	{paraskevop@ceid.upatras.gr}
	{}%https://orcid.org/...
	{}%(Optional) author-specific funding acknowledgements

\author{Christos {Zaroliagis}}
	{Computer Engineering and Informatics Department, University of Patras, Greece
	\and Computer Technology Institute and Press ``Diophantus'', Greece}
	{zaro@ceid.upatras.gr}
	{https://orcid.org/0000-0003-1425-5138}
	{}%(Optional) author-specific funding acknowledgements

\authorrunning{S. Kontogiannis, A. Paraskevopoulos, C. Zaroliagis}

\Copyright{Spyros Kontogiannis, Andreas Paraskevopoulos, Christos Zaroliagis} %TODO mandatory, please use full first names. LIPIcs license is "CC-BY";  http://creativecommons.org/licenses/by/3.0/
 %TODO mandatory: Please choose ACM 2012 classifications from https://dl.acm.org/ccs/ccs_flat.cfm 

 %TODO mandatory; please add comma-separated list of keywords

\category{} %optional, e.g. invited paper

\relatedversion{} %optional, e.g. full version hosted on arXiv, HAL, or other respository/website
\funding{
	This work was co-financed by the European Regional Development Fund of EU and Greek national funds
	through the Operational Program Competitiveness, Entrepreneurship and Innovation (call RESEARCH-CREATE-INNOVATE)
	under contract no. T2EDK-03472 (project ``i-Deliver'').
}%optional, to capture a funding statement, which applies to all authors. Please enter author specific funding statements as fifth argument of the \author macro.

\acknowledgements{}%optional

\nolinenumbers %uncomment to disable line numbering

\EventEditors{}
\EventNoEds{0}
\EventLongTitle{ArXiv Technical Report}
\EventShortTitle{ArXiv Report}
\EventAcronym{ArXiv}
\EventYear{2024}
\EventDate{}
\EventLocation{}
\EventLogo{}
\SeriesVolume{}
\ArticleNo{}
\usepackage{multirow}
\usepackage{graphicx}
\usepackage{array}
\usepackage{verbatim}
\usepackage{algpseudocode}
\usepackage{mathtools}
\usepackage[short]{optidef}

\usepackage{color, xcolor, colortbl,soul}

\newcommand{\Ind}  [1]{\mathbb{I}_{\left\{#1\right\}}}

\newcommand{\reals}{ \mathbb{R} }

\newcommand{\positivereals}{ \mathbb{R}_{\mbox{\tiny $>\!\! 0$}} }

\newcommand{\term}[1]{{\em #1}}

\newcommand{\AND}      {\wedge}

\newcommand{\OR}       {\vee}

\newcommand{\VRP}[1]		{\mathbb{VRP}{\scriptscriptstyle\mathrm{#1}}}
\newcommand{\VRPPDSTC}		{\VRP{PDSTC}}
\newcommand{\VRPPDSTCtd}	{\VRP{PDSTCtd}}

\newcommand{\Vstart}       	{\mathcal{V}^{start}}
\newcommand{\Vend}       	{\mathcal{V}^{end}}
\newcommand{\Vpickup}		{\mathcal{V}^{pic}}
\newcommand{\Vdelivery}		{\mathcal{V}^{del}}

\newcommand{\requestEarliestPickupTime} 	{t^{ep}}
\newcommand{\requestLatestDeliveryTime} 	{t^{ld}}
\newcommand{\requestReleaseTime} 	{t^{rel}}
\newcommand{\pickupServiceTime} 	{t^{psrv}}
\newcommand{\deliveryServiceTime} 	{t^{dsrv}}
\newcommand{\pickupPoint}  			{\chi^{pic}}
\newcommand{\deliveryPoint}  		{\chi^{del}}

\newcommand{\vertexStartTime}		{t^{start}}
\newcommand{\vertexEndTime}		{t^{end}}

\newcommand{\workerStartPoint}		{\chi^{start}}
\newcommand{\workerEndPoint}		{\chi^{end}}
\newcommand{\workerStartTime}		{t^{start}}
\newcommand{\workerEndTime}		{t^{end}}
\newcommand{\vehicleType}			{H}
\newcommand{\vehicleCapacity}		{Q}

\newcommand{\ALGORITHM}[1]			{\texttt{#1}}
\newcommand{\PLAININSERTION}[1]		{\ALGORITHM{Plain-Insertion}{#1}} 
\newcommand{\TDINSERTION}[1]		{\ALGORITHM{TD-Insertion}{#1}}
\newcommand{\TDPROPHET}[1]			{\ALGORITHM{TD-Prophet}{#1}}

\begin{document}

\maketitle

\begin{abstract} %\small\baselineskip=9pt 

The \term{Vehicle Routing Problem with pickups, deliveries and spatiotemporal service constraints} ($\VRPPDSTC$) is a quite challenging algorithmic problem that can be dealt with in either an offline or an online fashion. In this work, we focus on a generalization, called $\VRPPDSTCtd$, in which the travel-time metric is \emph{time-dependent}: the traversal-time per road segment (represented as a directed arc) is determined by some function of the departure-time from its tail towards its head. Time-dependence makes things much more complicated, even for the simpler problem of computing earliest-arrival-time paths which is a crucial subroutine to be solved (numerous times) by $\VRPPDSTCtd$ schedulers.
We propose two \emph{online} schedulers of requests to workers, one which is a time-dependent variant of the classical \ALGORITHM{Plain-Insertion} heuristic, and an extension of it trying to digest some sort of forecasts for future demands for service. We enrich these two online schedulers with two additional heuristics, 
	one targeting for distance-balanced assignments of work loads to the workers and 
	another that makes local-search-improvements to the produced solutions.
We conduct a careful experimental evaluation of the proposed algorithms on a real-world instance, with or without these heuristics, and compare their quality 
	with human-curated assignments provided by professional experts (human operators at actual pickup-and-delivery control centers), and also 
	with feasible solutions constructed from a relaxed MILP formulation of $\VRPPDSTCtd$, which is also introduced in this paper. 
Our findings are quite encouraging, demonstrating that the proposed algorithms produce solutions which 
	(i) are significant improvements over the human-curated assignments, and 
	(ii) have overall quality pretty close to that of the (extremely time-consuming) solutions provided by an exact solver for the MILP formulation.

\end{abstract}

%%%%%%%%%%%%%%%%%%%%%%%%%%%%%%%%%%%%%%%%%%%%%%%%%%%%%%%%%%%%%%%%%%%%%%%%%%%%%%%%%%
\section{Introduction}
\label{sec:introduction}

The vehicle routing problem with pickups, deliveries and spatiotemporal service constraints, $\VRPPDSTC$, concerns the utilization of a fleet of \emph{workers} (e.g., drivers, couriers, etc.) with their own work-shifts and capacitated vehicles, for the provision of one-to-one delivery services of \emph{commodities} (e.g., parcels, food, individuals, etc.) from their origins (\emph{pickup points}) to their destinations (\emph{delivery points}) within certain hard time-windows which are determined by earliest pickup-times and latest delivery-times per commodity. 
The primary goal is to have a maximum number of served commodity-delivery requests by the fleet of workers, respecting all spatiotemporal constraints (i.e., vehicle capacities, work shifts, and servicing time-windows), with a secondary objective that the workers commute in an underlying road network of a (typically large-scale) urban area in such a way that a specific aggregate service-cost function (e.g., total travel-time, total-distance of the entire fleet, etc.) is minimized.

An even more complicated generalization of the problem, $\VRPPDSTCtd$, considers instances in which the traversal-times of the road segments (which are represented as directed arcs), rather than being scalars, are \emph{time-dependent}, i.e., they are determined by given 
	%(typically FIFO-abiding, continuous and piecewiselinear)
arc-traversal-time functions of the departure-times from their tails towards their heads. Such a travel-time metric is typical when computing earliest-arrival-time paths for private vehicles commuting within road networks, but unfortunately makes the problem of computing earliest-arrival-time paths much harder 
(cf.~\cite{2022-Kontogiannis-et-al-axiomatic-approach} and references therein). Since this is a typical subroutine that must be used numerous times when solving an instance of $\VRPPDSTCtd$, it is clear that this generalization of the vehicle routing problem becomes even harder as well.  

The problem can be dealt with either offline, i.e., having at the solver's disposal the entire instance of delivery requests to be served and the fleet of workers, or online, i.e., when the requests for commodities to be delivered appear in real-time and/or the workers are activated at will. 
	% OFFLINE APPROACH
For $\VRPPDSTC$, typical approaches for the offline case such as the consideration of an appropriate mixed-integer linear programming formulation and the use of state-of-the-art MILP solvers, are well-known but also extremely demanding in computational resources, since the problem is NP-hard to solve. Unfortunately, for $\VRPPDSTCtd$ the situation becomes even more complicated, since there is no MILP formulation to solve (the travel-time metric is not constant but time-dependent). 

	% ONLINE APPROACH
Therefore, our focus is mainly on the efficient construction of suboptimal solutions in an online scenario where the work-shifts are predetermined and a priori known to the scheduler, but the requests are revealed in real-time and the scheduler has to always maintain a feasible solution for a maximal number of the active requests by the currently operational workers. As it is not obvious how classical constructive and improvement heuristics for $\VRP{}$ can be adapted when the service requests come in pickup-delivery pairs (one per served commodity), the literature for $\VRPPDSTC$ has mainly focused on simple online solvers, namely some well-known constructive heuristics such as the \ALGORITHM{Neighborhood} and the \ALGORITHM{Insertion} heuristics. 
	% INSERTION-heuristic
In particular, \ALGORITHM{Insertion} is a popular online algorithm, heavily used and experimented in the past, e.g., in~\cite{2017-Ceng-Xin-Chen,2022-Tong-et-al} for $\VRPPDSTC$, which simply constructs incrementally a feasible solution by allocating in a locally-optimal way each emergent request to one of the existing routes (creating a new route also being an option, provided there exist active workers still awaiting their first assignment) in such a way that the relative order of the already assigned requests remains intact and the incremental cost in the value of the objective function is minimized. Typically this heuristic requires cubic time, but there are also some quite efficient (even linear-time) implementations based on dynamic programming~\cite{2022-Tong-et-al}. Incorporating a time-dependent travel-time metric in such heuristics is already a challenge.
On the other hand, the adaptation of well-known exact polynomial-size MILP formulations for $\VRPPDSTC$ to $\VRPPDSTCtd$, to be fed to an offline solver, seems to be very hard because the point-to-point travel-times are now time-dependent variables rather than scalars.
	
In this work, we propose, implement, engineer, and experimentally evaluate two insertion-based algorithms for $\VRPPDSTCtd$: 
	The \TDINSERTION{} and the \TDPROPHET{}. 
Our implementation of \TDINSERTION{}, apart from the typical greedy criterion (the minimization of the additive cost for fitting a new request in the subtour of a worker) for accommodating an emergent commodity to some active worker, also considers an alternative local-optimization criterion, which essentially attempts to keep a rough balance in the aggregate lengths of the workers' subtours. This criterion was inspired by~\cite{2020-Bektas-Letchford} who observed, in the most elementary variant of $\VRP{C}$ only with vehicle capacities, that in optimal solutions some subtours correspond to much longer routes than others. Trying to avoid this kind of unfairness among the workers' actual commodity-servicing tasks, they proposed to compute the scores (i.e., marginal increases in route lengths) of the candidate pairs using the \emph{difference of squared costs} (the $\ell_2$-scoring criterion), rather than just the difference of the route costs (the $\ell_1$-scoring criterion). We implement and experimentally evaluate for $\VRPPDSTCtd$ the local-optimization analogue of the $\ell_2$-scoring criterion for our \TDINSERTION{} heuristic. As an alternative, we also try to hard-code fairness in the workers' subtour lengths when considering the classical $\ell_1$-scoring criterion, via an additional heuristic feature that we may opt to use in our scheduler, called the \ALGORITHM{Work Balancer} heuristic.

	Our \TDPROPHET{} algorithm was inspired by the \ALGORITHM{Prophet-Insertion} algorithm of \cite{2022-Tong-et-al} and works similarly with \TDINSERTION{}, but also tries to account for some sort of \emph{forecasts} for near-future requests and handles them exactly as the actual requests. Apart from the consideration of the time-dependent travel-time metric, another difference from the \ALGORITHM{Prophet-Insertion} algorithm of~\cite{2022-Tong-et-al} is that \TDPROPHET{} does not have the workers always on the move just because of predictions for the entire period (as \ALGORITHM{Prophet-Insertion} does in the static case); it just fits a small number of \emph{short-term} predictions (e.g., only within the next hour of operation) to their actual assignment of real requests that appear online to the system, and simply shortcuts the coverage of delivery points of those predictions that were not eventually verified in real-time at their pickup points.
	
Apart from implementing and engineering our online algorithms for $\VRPPDSTCtd$, we also evaluate the efficiency of a local-search improvement heuristic, namely, the repetitive \ALGORITHM{Relocation} of already assigned but not yet served routes right after handling a new (real or predicted) request, towards improving the solutions produced by the two algorithms.

	% RESPONSE TO REVIEWER#3 COMMENT ON MILP-RELAXATION
	As it would be too expensive to have an \emph{exact} mixed-integer linear programming (MILP) formulation for $\VRPPDSTCtd$\footnote{One could possibly consider a set-partitioning formulation using all feasible routes, but this would require exponentially many variables.}, we also propose a heuristic construction of some ``baseline'' solutions, using a \emph{relaxed} MILP formulation of polynomial size. This MILP considers a \emph{scalar} travel-time metric for interconnecting routes of service points of the requests which, rather than being just the average travel-times or the (optimistic) free-flow travel-times or the (pessimistic) full-congestion metric, are deduced by some ``educated'' estimations (scalars) of the actual time-dependent travel-times, depending on when these interconnecting routes are most likely to be used by any worker. Well-known MILP solvers are then used to create, within bounded execution time, a small set of solutions which are then tested for feasibility w.r.t. the temporal constraints, under the actual (time-dependent) travel-time metric. 
	This way we get some ``baseline solutions'' with which the solutions of our online algorithms are compared. Of course, even an optimal solution for the relaxed MILP is not necessarily an optimal solution for the time-dependent instance at hand, and its cost does not necessarily constitute some guaranteed lower-bound of the optimal cost. Of course, this MILP-based method is rather unrealistic for our online scenario, due to both the assumption of a priori knowing all delivery requests and its need for extremely demanding computational resources.

Finally, we conduct a thorough experimental evaluation of our online algorithms for $\VRPPDSTCtd$, with or without the heuristic improvements, on a real-world instance of food and supermarket delivery requests in an urban environment, which is fed with synthetic demand forecasts of varying accuracy. It is mentioned at this point that although high-quality demand forecasting is of paramount importance, it is \emph{not} the subject of the present work. This is why, for the purposes of our experimental evaluation only, we created synthetic forecasting data of varying accuracy. 
As a measure of comparison for our produced solutions, we use both the feasible solutions constructed by the relaxed MILP formulation and the actual solution that was determined by experienced human operators in our real-world dataset. Our results demonstrate a significant prevalence of both our online algorithms over the human-curated solution, up to 49\% in total length and travel time, and also the prevalence of \TDPROPHET{} over \TDINSERTION{}, up to 4\%, on finding better pickup-delivery scheduling solutions.

%%%%%%%%%%%%%%%%%%%%%%%%%%%%%%%%%%%%%%%%%%%%%%%%%%%%%%%%%%%%%%%%%%%%%%%%%%%%%%%%%%
\section{Problem Statement and Related Work}
\label{sec:problem-statement}

We are given a sequence
	$\mathcal{R} = \langle r_1,r_2,\ldots,r_{|\mathcal{R}|}\rangle$
of \emph{pickup-and-delivery} requests. Each request is a tuple 
	\(
		r = \left(
					\pickupPoint_r,
					\requestEarliestPickupTime_r,
					\pickupServiceTime_r,
					\deliveryPoint_r,
					\requestLatestDeliveryTime_r,
					\deliveryServiceTime_r,
					q_r,
					h_r
			\right) 
			\in \mathcal{R}
	\),
	where:
	%\item 
	$\requestEarliestPickupTime_r$ ($\requestLatestDeliveryTime_r$) is the \emph{earliest-pickup-time} (\emph{latest-delivery-time}) that a worker may receive (leave) the commodity from (at) the \emph{pickup point} $\pickupPoint_r$ (\emph{delivery point} $\deliveryPoint_r$), assuming that $\requestEarliestPickupTime_r < \requestLatestDeliveryTime_r$; 
	%\item 
	$\pickupServiceTime_r$ ($\deliveryServiceTime_r$) is the anticipated \emph{service-time} for the worker that is responsible for commodity $r$, at the corresponding (pickup or delivery) location ($\pickupPoint_r$ or $\deliveryPoint_r$);  
	%\item 
	$q_r$ is the \emph{volume/weight} of the commodity to be transferred, that is consumed from the corresponding vehicle's capacity;
	%\item 
	and $h_r\subseteq \mathcal{H}$ is the subset of \emph{eligible vehicle-types} for the good to be transferred (e.g., bicycle, motorcycle, car, with a cooler or heated box, etc.).
$\Vpickup = \{ (r,\pickupPoint_r): r\in \mathcal{R} \}$ and  $\Vdelivery = \{ (r,\deliveryPoint_r): r\in \mathcal{R} \}$ are the sets of pickup and delivery events, respectively, for all the active requests in $\mathcal{R}$. It is noted that, even if two requests $r\neq r'$ share some (geographical) service point, e.g., $\pickupPoint_r = \deliveryPoint_{r'}$ or $\pickupPoint_r = \pickupPoint_{r'}$, the pairs $(r,\pickupPoint_r),(r,\deliveryPoint_r), (r',\pickupPoint_{r'}), (r',\deliveryPoint_{r'})$ are distinct.

There is also a set $\mathcal{W} = \left\{ w_1, w_2, \ldots, w_{|\mathcal{W}|} \right\}$ of active workers (e.g., operational couriers during a work-shift), each of them represented by a tuple
	\(
		w = (
				\workerStartPoint_w,
				\workerStartTime_w,
				\workerEndPoint_w,
				\workerEndTime_w,
				\vehicleCapacity_w,
				\vehicleType_w)
	\)
	where:  
	%\item 
	$\workerStartPoint_w$ and $\workerStartTime_w$ ($\workerEndPoint_w$ and $\workerEndTime_w$) are the initial (final) \emph{location} and opening (closing) time, respectively, of $w$'s work-shift;
	%\item 
	$\vehicleType_w\in\mathcal{H}$ is the \emph{type} of the particular vehicle used by $w$ (e.g., bicycle, motorcycle, car, etc); 
	%\item 
	$\vehicleCapacity_w$ is the maximum \emph{volume/weight} of storage, for the vehicle used by $w$.
	$\Vstart = \{ (w,\workerStartPoint_w): w\in \mathcal{W} \}$ and  
	$\Vend = \{ (w,\workerEndPoint_w): w \in \mathcal{W} \}$ 
	are the sets of work-shift starting and finishing locations, for all the active workers. 

	Each worker $w\in\mathcal{W}$ may be assigned an arbitrary subset of requests $\mathcal{R}_w\subseteq \mathcal{R}$ which are eligible for them to serve. 
	The whole task for $w$ is represented as a sequence of all the corresponding pickup and delivery points for requests of $\mathcal{R}_w$, called his/her \emph{subtour}. 
	Then, $w$ is assumed to move within an urban area along earliest-arrival-time subpaths connecting consecutive points in the subtour, in order to serve them. 
	The area is represented by a directed graph $G = (V,E)$, whose arcs correspond to unidirectional road segments and vertices represent intersections and intermediate 
	points (corresponding to distinct postal addresses) of these road segments. Each arc $e=uv \in E$ comes with a scalar \emph{arc-length}, $\lambda[e]$, and a periodic 
	\emph{arc-travel-time} function $\tau_h[e](t) : [0,T) \mapsto \reals_{\geq 0}$ for evaluating the traversal-time of $e$ when using a vehicle of type $h \in \mathcal{H}$, 
	depending on the departure-time from $u$. 
	For succinctness in its representation, this function is assumed to be continuous and piecewise linear (pwl), represented as a constant-size sequence of breakpoints. 
	It is also assumed to satisfy the FIFO property, as is typical for individually moving private vehicles within road networks. The FIFO property implies that the 
	corresponding \emph{arc-arrival-time} function $a_h[e](t) = t + \tau_h[e](t)$ for $e$ when using $h$ is \emph{non-decreasing}. 
In a similar fashion, we inductively define the notions of travel-time and arrival-time functions for paths which are perceived as sequences of incident arcs: 
	For each $k\geq 0$, a path 
		$\pi = \langle e_1 = (i_0,i_1), \ldots, e_{k} = (i_{k-1},i_{k}) \rangle$ and an arc $e_{k+1} = (i_{k},i_{k+1})$, 
		the path  $\pi \oplus e_{k+1}$ is constructed by appending $e_{k+1}$ at the end of $\pi$. 
	It then holds that
		$\lambda[\pi \oplus e_{k+1}] = \lambda[\pi] + \lambda[e_{k+1}]$, 
		$ a_h[\pi \oplus e_{k+1}](t) =  a_h[\pi](t) + \tau_h[e_{k+1}](  a_h[\pi](t) )$, and
		$\tau_h[\pi \oplus e_{k+1}](t) =  a_h[\pi \oplus e_{k+1}](t) - t$. 
Furthermore, $\tau_h[o,d](t_o)$ denotes the minimum path-travel-time, when departing at time $t_o$ from $o\in V$ towards $d\in V$, using a vehicle of type $h$, and the earliest arrival-time at $d$ is denoted as $ a_h[o,d](t)=\tau_h[o,d](t)+t$. 
%The scalar value $ff_h[o,d]$ denotes the minimum \emph{free-flow} path-travel-time from $o$ to $d$ with a vehicle of type $h$. 
The scalar $\lambda[o,d]$ denotes the minimum path-length from $o$ to $d$.

A \term{feasible solution} for an instance of $\VRPPDSTCtd$ is described as a collection $\left\{ S_w: w \in \mathcal{W}\right\}$ of \emph{subtours} (i.e., sequences of service points for all the requests assigned to them), one per worker, such that each request belongs to at most one subtour and, along each subtour $S_w$, there is no violation of a temporal constraint or a vehicle capacity constraint as $w$ moves with his/her vehicle between consecutive service points along $S_w$ across interconnecting paths of the road graph $G$. 
The primary goal is to find a feasible solution that maximizes the number of served (i.e., assigned) requests, and a secondary goal is to minimize a global cost objective value (e.g., total travel-time or total-length, for all workers). 

\begin{figure}[tbh]
	\centerline{\includegraphics[width=0.3\paperwidth]{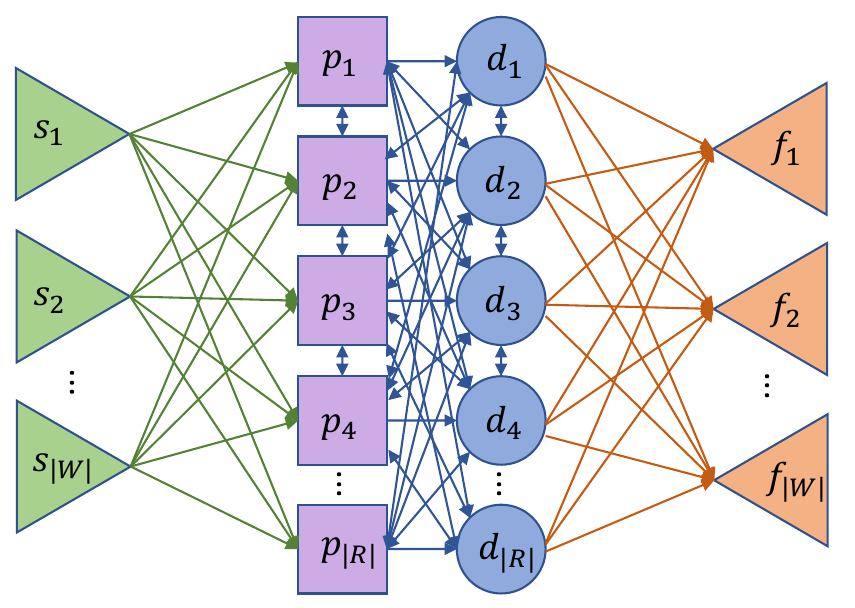}}
	\caption{\protect\label{fig:PD_GRAPH}%
		The pickup-and-delivery (PD) graph. 
	}%CAPTION
\end{figure}

For convenience, we consider a special graph, the \emph{pickup-and-delivery} (PD) graph $G_{PD} = (\mathcal{V},\mathcal{E})$ (cf. Figure~\ref{fig:PD_GRAPH}), whose node set $\mathcal{V}$ contains four subsets of nodes corresponding to distinct events: 
	The green and orange nodes correspond to workers-shift starting and ending events from $\Vstart$  and $\Vend$, respectively.
	The purple and blue nodes correspond to pickup and delivery events from  $\Vpickup$ and $\Vdelivery$, respectively. 
	As for the arc set $\mathcal{E}$, 
		nodes from $\Vstart$ are connected to all nodes in $\Vpickup$, 
		nodes from $\Vdelivery$ are connected to each node in $\Vend$, and
		(roughly) a complete subgraph is induced by $\Vpickup\cup\Vdelivery$, excluding only arcs from each delivery event to the pickup event of the same request, as they cannot be part of any solution. 
	
	For each $u\in \Vpickup\cup\Vdelivery$, $\rho(u) \in \mathcal{R}$ denotes the corresponding request. 
	For each $v \in \Vstart\cup\Vend$, $\gamma(v) \in \mathcal{W}$ denotes the corresponding worker for the work-shift $v$. 
	Within $G_{PD}$, each subtour $S_w = \langle v_0,v_1,\ldots,v_{k+1}\rangle$ can be seen as a (simple) path where $v_0=\workerStartPoint_w$, $v_{k+1}=\workerEndPoint_w$, and $\forall i\in\{1,2\ldots,k\}, v_i \in \Vpickup \cup \Vdelivery$ (service point for some request).

	%RESPONSE TO REVIEWER#3-L209
For each arc $uv\in\mathcal{E}$, there is 
	a minimum-length path $\pi^{\lambda}_{u,v}$ (and possibly suboptimal travel-time), and 
	a minimum-travel-time path $\pi^{\tau}_{u,v}(t_u)$, dependent on the departure-time $t_{u}$ (and possibly suboptimal length) 
in the underlying road graph $G$ connecting $u$ and $v$. 
Each subtour (i.e., simple path in $G_{PD}$) $S_w$ of a worker can then be translated within the road graph $G$ into a route by using
	either a distance-optimizing \emph{route} $\Pi^{\lambda}_w$ (prioritizing the usage of length-optimal interconnecting paths), or
	a travel-time-optimizing route $\Pi^{\tau}_w(t_w^{start})$ (prioritizing the usage of length-optimal interconnecting paths), 
that interconnects all the consecutive points in $S_w$. 
For some route $\Pi_w = \left( v_0=\workerStartPoint_w, v_1, \ldots, v_{k+1}=\workerEndPoint_w \right)$ for worker $w\in\mathcal{W}$, we denote by $\Pi_w^{u:v}$ the subroute starting at node $u$ and ending at node $v$. For a given departure-time $t$, we associate each node $v_i \in \Pi_w$ with the following labels: 
	\begin{enumerate}

	\item
		\emph{arrival-time}: $a(v_i) = a[\Pi_w^{v_0:v_i}](t)$; 

	\item
		\emph{earliest departure-time}:
			\(
				d(v_i) = \left\{\begin{array}{rl}
					\max\{ a(v_i) , \requestEarliestPickupTime_{\rho(v_i)} \} + \pickupServiceTime_{\rho(v_i)},
					& v_i \in \Vpickup
					\\ 
					a(v_i) + \deliveryServiceTime_{\rho(v_i)},
					& v_i \in \Vdelivery
					\\ 
					a(v_i), 												
					& \mbox{otherwise}
				\end{array}\right.
			\)		 
	\item
		\emph{waiting-time}:
				$b(v_i) = \max\{ \requestEarliestPickupTime_{\rho(v_i)} - a(v_i), 0 \}$ for $v_i \in \Vpickup$, and $b(v_i)=0$ otherwise; and 
	\item
		\emph{current-load}:
			\( 
				C(v_0)=0;~
					\forall i\geq 1, 
					C(v_i) = 
						\left\{\begin{array}{rl}
							C(v_{i-1}) +  q_{\rho(v_i)},	& v_{i} \in \Vpickup
							\\
							C(v_{i-1}) - q_{\rho(v_i)}, 	& v_{i} \in \Vdelivery
							\\C(v_{i-1}), 	& \mbox{otherwise}
						\end{array}\right.
			\)			

	\end{enumerate}

When considering to insert a new service point $u$ in $S_w$ right after some existing point $v_i$, all the subsequent subpaths interconnecting consecutive nodes of $S_w$ after $v_i$ must be recomputed, to account for the updated departure and arrival times along $\Pi_w$. Due to the time-dependent nature of the travel-time metric, this is a non-trivial task to execute, prior to assessing the effectiveness of positioning the new event at a particular place within $\Pi_w$. 

An instance of  $\VRPPDSTCtd{}$ is represented by 
	a directed graph $G=(V, E)$, 
	scalar arc-lengths $\lambda: E \mapsto \positivereals$, 
	(periodic, continuous and piecewise-linear) arc-travel-time functions $(\tau_h[e]:[0,T)\mapsto\positivereals)_{h\in H, e\in E}$,
	a sequence of requests $\mathcal{R}$, and a set of workers $\mathcal{W}$.
As previously mentioned, we also construct the auxiliary graph $G_{PD}$. 
Because there may be nodes in $\mathcal{V}$ whose geo-locations do not coincide with vertices in $V$, some road-pedestrian connections are added between them by finding the nearest-neighbor pairs $(x,v)$, for each $x\in V$ and $v \in \mathcal{V}$. The nearest-point search is done efficiently using an R-tree~\cite{1984-Guttman}. 
We also conduct sequentially shortest-path-tree computations for any involved vehicle type, to provide a set of (one-to-many) minimum-length paths and minimum-travel-time paths among geo-locations for elements of $V\cup \mathcal{V}$. For the (scalar) length metric we simply employ executions of Dijkstra’s algorithm~\cite{Dijkstra}. As for the time-dependent travel-time metric, the earliest-arrival-time computations are efficiently performed ``on the fly'', using the query algorithm \ALGORITHM{CFCA} of the CFLAT oracle~\cite{CFLAT} for time-dependent shortest paths, exactly when an arc in the PD graph $G_{PD}$ is to be used by some worker. This is something that can be done efficiently by an online algorithm that only tries to fit into an existing solution a single new delivery request. On the contrary, the consideration of time-dependent travel-times renders impossible the construction of an exact MILP formulation; therefore, even the time-consuming construction of an optimal solution via MILP solvers becomes quite more challenging in this case. 
Section~\ref{sec:MILP-relaxation-of-VRPPDSTCtd} in the appendix provides an approximate MILP formulation for $\VRPPDSTCtd{}$ that considers some carefully selected scalar travel-time values for entire paths (rather than just arcs), exactly when they could be possibly used by some (any) feasible solution. 

%%%%%%%%%%%%%%%%%%%%%%%%%%%%%%%%%%%%%%%%%%%%%%%%%%%%%%%%%%%%%%%%%%%%%%%%%%%%%%%%%%
\section{Insertion-based Schedulers for $\VRPPDSTCtd{}$}
\label{sec:VRPPD-algorithms}

The purpose of an insertion heuristic is to assign each request $r$ to a worker $w$ in a cost-optimal way, so that the new subtour $S'_w$ (after adding the two service points of $r$) maintains the same relative order for the service points in $S_w$. The main reasons for such a requirement are simplicity and computational efficiency, since the consideration of all possible subtours for $R'_w = R_w\cup\{r\}$ would require the examination of an exponential number of subtours~\cite{2019-Pan-et-al}.

\begin{definition}[Insertion-Based Heuristics]
	\protect\label{dfn:insertion-heuristic}
	Given a collection of subtours $S_w$ for serving the subsets $R_w$ of requests assigned to each operational worker $w\in\mathcal{W}$, and a new request $r$,  an \emph{Insertion-Based heuristic} determines for each $w\in \mathcal{W}$ a candidate subtour $S'_w$ for $R'_w = R_w \cup \{r\}$, which achieves a minimum increase in $w$'s contribution to some global-objective value, and leaves intact the relative order of the service points already in $S_w$. Eventually, $r$ is assigned to the worker achieving the minimum increase, among all workers.
\end{definition}

The \PLAININSERTION{} algorithm is well known in the literature of $\VRP{}$-related problems. A na\"ive  implementation of such a heuristic would require quadratic, or even cubic computational time. 
A linear-time implementation of \PLAININSERTION{} for $\VRPPDSTC$ was recently proposed~\cite{2022-Tong-et-al}, which is based on a preprocessing step and on dynamic programming for computing the workers' scores (i.e., the marginal increases in cost if they were assigned the new request), for a scalar travel-time metric. 
We introduce in this section a variant of \PLAININSERTION{}, called \TDINSERTION{} for $\VRPPDSTC$, which follows the main idea of the preprocessing of workers' paths in~\cite{2022-Tong-et-al}, so as to achieve early pruning of infeasible solutions, but with some major modifications so as to deal with the \emph{time-dependent} travel-time metric and the consideration of earliest pickup-times for each request. The primary objective is to maximize the number of assigned requests. As a secondary objective, our algorithm considers two alternatives, as was previously explained: either the sum (i.e., $\ell_1$-norm), or the sum-of-squares (i.e., $\ell_2$-norm) of the workers' costs (distances, or travel-times).
	The $\ell_2$-scoring criterion was inspired by~\cite{2020-Bektas-Letchford}, as an indirect means of \emph{inducing} more balanced allocations of requests to the workers. 
	We then incorporate in \TDINSERTION{} the heuristic \ALGORITHM{Workload Balancer} (WB), which \emph{enforces} some balance among the workers' assignments. We also consider a local-search improvement heuristic which post-processes the solutions provided by \TDINSERTION{}, exploring  among single-request relocation attempts for better solutions.
	Finally, we introduce \TDPROPHET{}, a variant of \TDINSERTION{} that, apart from actual requests, also includes in the produced subtours some short-term forecasts for future requests.

\subsection*{(a) Description of \TDINSERTION{}}
\label{sec:td-insertion-description}
%\vspace*{-10pt}
%\paragraph*{(a) Description of \TDINSERTION{}.}
We denote as \TDINSERTION{$_{\kappa,\nu}$} the variant of $\TDINSERTION{}$ that assumes a cost metric $\kappa\in\{\tau,\lambda\}$ ($\tau$ for travel-times and $\lambda$ for distances) and a norm $\ell_{\nu}\in\{\ell_1,\ell_2\}$ for assessing the scores of candidate insertion pairs of each new request. 
It is assumed inductively that: 
	(i) each worker $w\in\mathcal{W}$ has already been assigned a subset of requests $R_w$, to be served according to the subtour (i.e., sequence) $S_w$ of the corresponding service points; 
	(ii) $S_w$ has already been translated into some particular route $\Pi^{\kappa}_w$, depending on the particular cost metric $\kappa\in\{\tau,\lambda\}$ that we consider as primary. 
It should be noted at this point that $\Pi^{\tau}_w$ is indeed a route of minimum-travel-time interconnecting subtours. On the other hand, $\Pi^{\lambda}_w$ is not necessarily a route of minimum-length interconnecting subtours. In particular, some minimum-travel-time interconnecting paths may also have been used in $\Pi^{\lambda}_w$ for some pairs of consecutive points in $S_w$, only as contingency interconnecting routes for the case that the insertion of a service point with length-optimal interconnecting paths has lead to a violation of some temporal constraint of the subsequent service points. More about this issue is discussed in Subsection \ref{sec:translating-assignments-to-routes-under-distance-metric}.
	
	Let $a(v_i)$ and $d(v_i)$ denote the arrival-time at $v_i$ and the departure-time from $v_i$, respectively, as $w$ moves along $\Pi^{cost}_w$. 
In a nutshell, the steps of \TDINSERTION{} are the following: 
For each new request $r$, at release-time $\requestReleaseTime_r$, we test the insertion of $r$ within the subtour
	$S_w=\langle v_0 = \workerStartPoint_w, v_1,v_2,\ldots,v_{|S_w|-1}=\workerEndPoint_w \rangle$ 
by iteratively placing the pickup node $\pickupPoint_r$ right after position $i\in\{0\ldots,|S_w|-2\}$ and the delivery node $\deliveryPoint_r$ right after position $j\in\{i,\ldots,|S_w|-2\}$. 
We also require that $\requestReleaseTime_r \leq d(v_{i+1})$, i.e., $r$ cannot precede a service point whose departure time is already before $r$'s release time. This insertion would result in the expanded subtour
	\(
		S'_{w} = 
			\langle v_0, v_1,\ldots,v_i,\pickupPoint_r,v_{i+1},\ldots,v_j,\deliveryPoint_r,v_{j+1},\ldots,v_{|S_{w}|-1}\rangle
	\)		
and the corresponding route $\Pi'_w$ from $\Pi^{cost}_w$ with the appropriate cost-optimal interconnecting paths. 
For each pair $(i,j)$ of candidate positions for the service points of $r$, a feasibility check of the spatial- and (time-dependent) temporal-constraints is performed along the suffix-subroute of $\Pi'_w$ starting at node $v_i$. If all these service points are still feasible, then a marginal-increase value $Score(\Pi^{cost}_w, i, j, r)$ is computed, to quantify the impact on $w$'s servicing cost of accepting the candidate positions $(i,j)$ of $S_w$ for serving $r$. In case of infeasibility, when $cost = \tau$ the candidate pair $(i,j)$ is immediately rejected. 
When $cost=\lambda$, we alternatively construct the route $\Pi''_w$ from $\Pi^{\lambda}_w$ with the appropriate travel-time-optimal interconnecting paths (only for the detours of $r$'s service points). 
We provide now a detailed description of exactly how this is done.

\subsubsection*{(a.i) Preprocessing check-constraint indicators for candidate insertions} 
%\vspace*{-10pt}
%\subparagraph*{(a.i) Preprocessing check-constraint indicators for candidate insertions.} 
	%
As in~\cite{2022-Tong-et-al}, given the $(cost,norm)$ pair that we consider, we use two check-constraint indicators for the service nodes of $S_w$, as $w$ moves along the corresponding route $\Pi_w$ (for simplicity, we slightly abuse notation by skipping the metric-dependent exponent, and the worker's shift-start-time): 
	\begin{itemize}

	\item 
		$slack(v_i)$ is the \emph{maximum tolerable time} for inserting a detour between $v_i$ and $v_{i+1}$, without violating any of the (temporal) latest-delivery-times for nodes in $\Pi_w^{v_{i+1}:v_{|S_w|-1}}$.

	\item 
		$ddl(v_i)$ is an upper bound on the ultimate arrival-time at $v_i$ so that neither the deadline for serving the request $\rho(v_i)$, nor the work-shift end of the carrying worker are violated. 

	\end{itemize}
For each arc $e = v_i v_{i+1} \in S_w\cap\mathcal{E}$, the travel-time $\tau[e](t)$, of the path $\pi_e=\langle v_i,v_{i+1} \rangle$ in $G$ associated with $e$, can be either increased or decreased as a function of the departure-time $t$ from $v_i$. Nevertheless, due to the FIFO property, the arrival-time function is non-decreasing: $\forall t< t',~ a[e](t)\leq a[e](t')$. Inserting $\pickupPoint_r$ between $v_i$ and $v_{i+1}$ will give a new arrival-time $a'(v_{i+1}) \geq a(v_{i+1})$. 
Therefore, the current arrival-time value $a(v_{i+1})$ is a lower-bound whereas $slack(v_i)$ is an upper bound on the arrival time at $v_{i+1}$, and their difference is an upper bound for the delay that may occur (due to some detours for adding new service points) between $v_i$ and $v_{i+1}$. 
In order to incorporate earliest pickup-times in the $slack(v_i)$ and $ddl(v_i)$ indicators, the following dynamic-programming approach is adopted, as we move backwards along $\Pi_w$, from the end $v_{|S_w|-1} = \workerEndPoint_w$ towards $v_{i}$:
	\begin{enumerate}
	
	\item 
		$ddl(v_i)$ values:
		\begin{itemize}

			\item for the work-shift end node, 
				$ddl(\workerEndPoint_w) = \workerEndTime_w$; 

			\item for a commodity-delivery node $v_i\in\Vdelivery\cap S_w$, 
				$ddl(v_i) = \requestLatestDeliveryTime_{\rho(v_i)}$; 

			\item for a commodity-pickup node $v_i \in \Vpickup\cap S_w: v_j \in \Vdelivery \AND r=\rho(v_i)=\rho(v_j)$, 
				\[
					ddl(v_i) = {\requestLatestDeliveryTime_r} 
					- \sum_{i \leq k < j} \left[ (a(v_{k+1})- d(v_k)) + t_k - b(v_k) \right]
				\] 
				where 
					\(
						t_k = \left\{\begin{array}{rl}
										\pickupServiceTime_{\rho(v_k)}, 	& v_k \in \Vpickup
										\\
										\deliveryServiceTime_{\rho(v_k)}, 	& v_k \in \Vdelivery
										\\
										0, 									& \mbox{otherwise}
							\end{array}\right.
					\)

		\end{itemize}
		 
	\item
		$slack(v_i)$ values:
		\(
			slack(v_k)= \min\left\{~ ddl(v_{k+1}) - a(v_{k+1}), slack(v_{k+1}) + b(v_{k+1}) ~\right\}
		\)
		for $k$ ranging from  $|S_w|-1$ down to $i$.  
		
		Recall that $b(v)$ represents the required waiting-time at a service node $v$, or just the resulting idle-time (only at pickup nodes), when $r$ imposes an earliest pickup-time $\requestEarliestPickupTime_r$.
		
	\end{enumerate}	

	It should be noted that, compared to a brute-force implementation of \ALGORITHM{Plain-Insertion}, the exploitation of these two auxiliary variables by our online algorithms allows the pruning without checking of many insertion-candidates, which has lead to a significant improvement of our implementations' execution times by at least 60\%. 
	
\subsubsection*{(a.ii) Efficient rejection of infeasible candidate insertions}
%\vspace*{-10pt}
%\subparagraph{(a.ii) Efficient rejection of infeasible candidate insertions.}
	%
The procedure is as follows: 
	Assume for a new request $r$ and a subtour $S_w$ that we consider for insertion the candidate pair $(i,j)$, for $0\leq i\leq j\leq |S_w|-2$.
	If 		\(
				[% I^{\tau} := 
					a'(v_{i+1}) - a(v_{i+1}) > slack(v_i) 	]
				\OR 
				[ 	C(v_{i}) + q_r > \vehicleCapacity_w 	]
			\)	 
			i.e., the resulting increase on the arrival-time at $v_{i+1}$ exceeds the slack of $v_{i}$,
			or the resulting vehicle-load after picking up $r$ at $v_{i}$ causes a violation in the vehicle capacity,
	then 	the candidate pair $(i,j)$ can be safely rejected. 
	Moreover, 
	if 		either of these two types of violation constraints occurs and $i<j$, 
			i.e., only the pickup node of $r$ is checked for insertion right after $v_i$, 
	then 	all the candidate pairs $(i,m) : i\leq m\leq j$ can be safely rejected. 

It should be noted at this point that, since the \emph{slack times} are only upper-bounds, even if the above checks are passed, we still need to check for potential violations in latest-delivery-times of requests or in the work-shift end-time along the suffix of the new subtour $S'_w$ that we create, under the time-dependent travel-time metric. 
Therefore, the time complexity to obtain all the feasible insertion-pairs for a new request along $S_w$ is $O(|S_w|^2)$, due to the unavoidable time-dependent travel-time updates when checking for these potential violations.

\subsubsection*{(a.iii) Computation of scores for feasible candidate insertion-pairs per route $\Pi_w$}
%\vspace*{-10pt}
%\subparagraph{(a.iii) Computation of scores for all feasible candidate insertion-pairs per route $\Pi_w$.}
	%
For a new request $r$, fix an arbitrary worker $w$ whose vehicle-type is eligible for $r$: $H_w \in h_r$. 
Recall that we consider some arc-cost metric $\kappa \in \{ \tau,  \lambda \}$ (i.e., traversal-times, or distances) for all the arcs in the auxiliary PD graph. As for the assessment of the scores for candidate insertion pairs, as already mentioned, we consider that it is specified by the norm $\ell_{\nu} \in \{\ell_1,\ell_2\}$.
For example, \TDINSERTION{$_{\tau,\ell_1}$} uses the travel-times cost metric for the routes and assesses the overhead of each candidate path according to the $\ell_1$ norm, whereas \TDINSERTION{$_{\lambda,\ell_2}$} uses the arc-lengths metric for the routes and the overhead of each candidate path according to the $\ell_2$ norm. 
For \TDINSERTION{} we need to compute one of the following path-costs for worker $w$'s subroute $\Pi_w$: 
	\begin{itemize}
	
	\item For distance-metric: 		$Cost_{\lambda}(\Pi_w) = \sum_{e=uv\in\Pi_w} (\lambda_e)$.
	 
	\item For travel-times metric: 	$Cost_{\tau}(\Pi_w) =\sum_{e=uv\in\Pi_w} [ a(v) - d(u) ]$.

	\end{itemize}
Given $\Pi_w$ and a particular candidate insertion pair $(i,j)$ for a new request $r$, $\Pi_w(i,j,r)$ is the resultant candidate subroute from $\Pi_w$ in which $\pickupPoint_r$ is positioned right after $v_i$ and $\deliveryPoint_r$ is positioned right after $v_j$ (and after $\pickupPoint_r$, in case that $i=j$). 
The \emph{scores} (i.e., marginal costs) of this subroute are calculated as follows, for $\kappa \in \{ \lambda, \tau\}$ and $\nu \in \{1,2\}$:
	\[
		Score_{\kappa,\ell_\nu}( \Pi_w,i,j,r ) 
		= 	\left\{\begin{array}{cl}
				[ Cost_{\kappa}( \Pi_w(i,j,r) ) ]^{\nu} - [ Cost_{\kappa}( \Pi_w )]^{\nu}, & \mbox{if $\Pi_w(i,j,r)$ is feasible}
				\\
				\infty, 															& \mbox{if $\Pi_w(i,j,r)$ is infeasible}
			\end{array}\right.		
	\] 
The score of $w$ for hosting $r$ is then 
	$Score_{\kappa,\ell_{\nu}}( w , r ) = \min_{0\leq i\leq j\leq |S_w|-2}Score_{\kappa,\ell_{\nu}}( \Pi_w,i,j,r )$.
Eventually, $r$ is assigned to a worker $\hat{w}$ of minimum score:
	$\hat{w} \in\arg\min_{w\in\mathcal{W}} Score_{\kappa,\ell_{\nu}}( w,r )$.

Note that during step (a.ii), for each feasible pair $(i,j)$, the score can be computed in parallel with the constraint-checking process. When a feasible pair $(i,j)$ is verified, its (finite) score is compared to the minimum score discovered so far. Then, $(i,j)$ is rejected immediately when the calculation of its score already gives a value larger than the current minimum score. 

A particular attention should be given when working with distances: 
	If some $(i,j)$ leads to an eventually \emph{infeasible route} $\Pi'_w := \Pi_w(w,i,j,r)$, because of violations of temporal constraints, 
	then we repeat the process using travel-time-optimal (instead of distance-optimal) interconnecting paths for pairs of consecutive service points along $\Pi_w$ (corresponding to arcs in the PD graph). This approach guarantees that, if there is any feasible solution at all for the new request, then the request will at least be allocated to some subroute $\Pi''_w$ of finite score, even if having to use distance-suboptimal interconnecting paths.
For $\TDINSERTION{}$ we considered two distinct contingency plans when facing such infeasibilities with the distance metric:
	% (1) 
	(i) either construct $\Pi''_w$ separately for each $(i,j)$ whose $\Pi'_w$ is time-infeasible, interconnecting with distance-suboptimal subtours $r$'s service points within $\Pi_w$, or
	% (2) 
	(ii) recompute from scratch an assignment for $r$, under the travel-time objective this time, but only when all the candidate pairs under the distance metric provided temporally infeasible routes. 
Eventually our decision for the experimental evaluation of $\TDINSERTION{}$ was to adopt the former contingency plan, as it adopts the travel-time metric not for each and every candidate pair but only for the problematic detours.

\subsection*{(b) Workload Balancer Heuristic (WB)}
%\vspace*{-10pt}
%\paragraph*{(b) Workload Balancer (WB) Heuristic}
When running the experiments, it was observed that the optimal solutions provided by \TDINSERTION{} involved only a few workers that shouldered the majority of the requests, while the rest of the workers did much less work, or were even not assigned any request at all. 
Towards providing more fair assignments for all the operational workers, we consider a threshold $\theta \geq 1$ and a penalty factor $\mu \geq 1$ and we introduce a bias for new requests in favor of workers with lighter (by means of traveled distance) workloads, even though some other workers might serve them with smaller marginal service costs. 
This bias is achieved by our \ALGORITHM{Workload Balancer} (WB) heuristic, which considers a slightly different scoring step for \TDINSERTION{} for determining the winning worker per new request. 
In particular, upon the release of a new request $r$, let 
	${\mathcal{W}}_o$ be the set of the currently operational workers, and 
	$Cost_{\kappa,\nu}(\Pi_w)$
be the cost of some worker $w \in \mathcal{W}_o$ for a given metric $\kappa \in \{\tau,\lambda\}$ and objective $\nu\in\{\ell_1,\ell_2\}$.
The total cost of the current solution (before serving $r$) is
	$Cost_{\kappa,\nu}(\mathcal{W}_o) = \sum_{w\in\mathcal{W}_o} Cost_{\kappa,\nu}(\Pi_w)$. 
Then, each operational worker $w\in\mathcal{W}_o$ whose subroute-cost exceeds the average subroute-length in $\mathcal{W}_o$ by more than $\theta$, gets a penalized score by a multiplicative factor $\mu > 0$: 
	$\forall w\in\mathcal{W}_o,~\forall \kappa\in\{\tau,\lambda\},~\forall \nu\in\{\ell_1,\ell_2\}$, 
	\[
			Score_{\kappa,\nu}^{wb}( \Pi_w,i,j,r ) 
			:= \left(1 + \mu\cdot\Ind{Cost_{\lambda,\ell_1}(\Pi_w) > \theta \cdot \frac{Cost_{\lambda,\ell_1}(\mathcal{W}_o)}{ |\mathcal{W}_o|}}\right) \cdot Score_{\kappa,\nu}( \Pi_w,i,j,r )
	\]

\subsection*{(c) Request Relocation Improvement Heuristic (RR)}
%\vspace*{-10pt}
%\paragraph*{(c) Request Relocation Improvement Heuristic (RR)}
A weakness of insertion-based heuristics is that they forbid changes in the assignment and the relative service order of the active requests, except for the new request.
Towards amplifying this drawback, as in~\cite{holThoLew2012}, we introduce the \ALGORITHM{Request-Relocation Improvement} (RR) heuristic, which conducts a sequence of local-search improvement attempts to the current solution as follows: 
For each $w\in \mathcal{W}_o$, and each $r\in\mathcal{R}_w$ that has \emph{not} been picked up yet by $w$, $\pickupPoint_r\in \Vpickup$ and $\deliveryPoint_r \in \Vdelivery$ are removed from $\Pi_w$, making the appropriate shortcutting to $\Pi_w$ so as to be a feasible subroute for $\mathcal{R}_w\setminus\{r\}$.
By the FIFO property, this may cause no violation of a spatiotemporal constraint in $\Pi_w$ and does not affect the routes of other workers. 
Consequently, $r$ is relocated by \TDINSERTION{}, either at a better position within $\Pi_w$ or within the route of another operational worker.

\subsection*{(d) Digesting Demand Forcasts with \TDPROPHET{}}
\label{sec:lazy-prophet}
%\vspace*{-10pt}
%\paragraph*{(d) \TDPROPHET{}: The Variant of \TDINSERTION{} that Digests Demand Forcasts}

In the typical online scenario, the request sequence $\mathcal{R}$ is initially unknown and is gradually revealed (per request) to the request scheduler. This knowledge gap is an important drawback for preparing a better and more organized scheduling plan. Apart from the higher risk of adopting suboptimal assignments, another significant burden is the necessity of the workers making large detours to serve newly revealed requests. 
Inspired by the \ALGORITHM{Prophet-Insertion} scheduler in~\cite{2022-Tong-et-al}, we introduce here a variant of \TDINSERTION{}, called the \TDPROPHET{}, which takes into account some sort of \emph{short-term forecasts} for future requests and deals with them exactly as (virtual) requests with pickup/delivery points and spatiotemporal constraints. These virtual requests are a priori scheduled in the front of the request sequence to be handled by the scheduler, so as to be assigned to (initially idle) operational workers. 
This assignment is done using \TDINSERTION{}. Consequently, their spatiotemporal constraints are deactivated (i.e., $q_r=0, \requestLatestDeliveryTime_r = \infty$), so as not to cause unnecessary infeasibilities for the workers' subroutes. 
The major difference of \TDPROPHET{} from \ALGORITHM{Prophet-Insertion} in~\cite{2022-Tong-et-al}, apart of handling time-dependent travel-times, is that, after determining the assignment of the predictions to the workers, the  delivery nodes of predictions of low appearance probability (below 80\%) are removed (to better deal with any forecast inaccuracy). Also, when \TDINSERTION{} decides to place the service nodes of a new request, say, at positions $i$ and $j$ respectively, any virtual pickup node (corresponding to a forecast) with low appearance probability (below 80\%) between $i$ and $j$ is simply ignored.
It should be noted that the demand-forecasting task is beyond the scope of this work and we consider this information to be provided as input to \TDPROPHET{}. Nevertheless, 
	%Section~B in the full version of the paper~\cite{2024-Kontogiannis-Paraskevopoulos-Zaroliagis-ARXIV}
	Section~\ref{sec:TDPROPHET-demand-forecasting} in the Appendix 
describes exactly how this forecasting task is \emph{simulated} for the real-world data set that we use for the needs of our experimentation.

%%%%%%%%%%%%%%%%%%%%%%%%%%%%%%%%%%%%%%%%%%%%%%%%%%%%%%%%%%%%%%%%%%%%%%%%%%%%%%%%%%
\section{Experimental Evaluation}
\label{sec:experimental-evaluation}

We evaluated our algorithms using a real-world data set with records of pickup-and-delivery food and shopping orders during 3 consecutive working days, at the midium-sized city of Ptolemaida, Greece. In our experiments we assess the performance of our online schedulers for the actual request-sequence against two baseline solutions: 
	%
	%\begin{enumerate}
	%	
	%\item
	(1) \emph{human-curated solutions} provided by operators in the control room of a middleware platform mastering the service of food-order and shopping-delivery requests in Ptolemaida; and 
	%
	%\item
	(2) optimal solutions to the \emph{MILP formulation for a relaxation} (a carefully constructed instance of $\VRPPDSTC$) of the actual instance of $\VRPPDSTCtd$ (the detailed description of this relaxation, the proposed MILP formulation and the adopted solution method, are provided in Section~\ref{sec:MILP-relaxation-of-VRPPDSTCtd} of the Appendix).
		
	%\end{enumerate}	

\subsection*{(a) Experimental Setup}
\label{sec:experimental-setup}
%\vspace*{-10pt}
%\paragraph*{(a) Experimental Setup}
	% Programming environment...
The algorithms are in C++ (GNU GCC v.11.3.0).
	% Machine details...
The experiments were conducted on an AMD EPYC 7552 48-Core 2.2GHz Processor with $256$GB RAM and Ubuntu (22.04 LTS).

\subsection*{(b) Experimental Dataset}
\label{sec:experimental-datasets}
%\vspace*{-10pt}
%\paragraph*{(b) Experimental Dataset}

The dataset contains a pair $(\mathcal{W},\mathcal{R})$ of a worker set and a request sequence, with actual pickup/delivery-times for the requests and work-shift intervals for the workers, within a period from Monday, July 3 2023 to Wednesday, July 5 2023, in the city of Ptolemaida in Northern Greece. All workers involved in this particular data set have used a single type of vehicle (scooters with a fixed-size storage). 
The human-curated service subtours were decided in real-time by well-experienced operators at the control center of a middleware platform providing couriers to food and shopping enterprises. The actual (GPS-recorded) service routes of the couriers were extracted from the pilot-phase event-logging database in the framework of a research project in which our group participated in the past~\cite{i-deliver}. These routes are already of high quality, since they were based on the long-term experience of the human operators, especially in the medium-size of the operational area (Ptolemaida). 
The construction of the road graph $G$ was based on an OpenStreetMap dataset for Greece's road network~\cite{osmfds}. The travel-time metric is provided by the OpenStreetMap service and the request-demand predictions were provided as input. $G$ contains $|V| = 2547$ nodes and $|E| = 9514$ arcs.
In the real data set some spatiotemporal restrictions were missing. In order to carry out a more realistic experimental evaluation, we adopted the following constraint scenario: 
		Each request was assumed to have one-unit load and a duration of $40min$ between the latest-delivery-time and and the earliest-pickup-time, and service times of $1.5min$;
		and each worker uses a vehicle with a total capacity of 3 units. I.e., 
		\(	
		 	\forall r\in\mathcal{R}~\left(~q_r = 1 \AND \requestLatestDeliveryTime_r = \requestEarliestPickupTime_r + 40min
		 	\AND
			\pickupServiceTime_r = \deliveryServiceTime_r = 1.5min~\right)
		 	\AND
			\forall w\in\mathcal{W}~\left(~\vehicleCapacity_w = 3 ~\right)\,.
		\)	
	
\subsection*{(c) Analysis of Experimental Results}
\label{sec:experimental-analysis}
%\vspace*{-10pt}
%\paragraph*{(c) Analysis of Experimental Results}

We executed three experiments, one per working day (Mon,Tue,Wed). Our online algorithms created the full sequences of worker subtours per day, starting from initially empty subtours. 
We experimented, exactly on the same instances, for an online algorithm
	\(
		alg \in 	\{		\TDINSERTION{^{heur}_{\kappa,\nu}}, 
						\TDPROPHET{^{heur}_{\kappa,\nu}} : 
							heur\in\{~ \{~\} , \{wb\} , \{rr\},\{wb,rr\} ~\},
							\kappa \in \{ \tau, \lambda \},
							\nu \in \{ \ell_1,\ell_2 \}
				\}
	\)
	where $heur$ indicates whether specific heuristics are activated, $\kappa$ determines the cost metric and $\nu$ specifies the type of the global objective.
	
Each variant of \TDPROPHET{$_{\kappa,\nu}$} works as follows: first we appended at the beginning of the request sequence a subset of predictions for virtual requests, which were then assigned to workers with \TDINSERTION{$^{wb}_{\kappa,\nu}$}. The remaining sequence (of the real requests) were handled then sequentially, exactly as they appeared, by \TDINSERTION{$_{\kappa,\nu}$}. Each real request was assumed to be visible to the scheduler only after its release time. 
Upon the release of a new (real) request $r\in\mathcal{R}$ at time $\requestReleaseTime_r$, each variant of our algorithms executes the following substeps: 
	$r\in\mathcal{R}$ is first assigned to a ``moving'' worker $w\in\mathcal{W}$ with the minimum score value, w.r.t. the objective function. 
	Then, a detour event takes place, if the worker is instructed to change destination. 
	Consequently, $w$'s route is expanded by adding the service nodes of the new request. 
	Finally, the new time-dependent interconnecting paths are computed, to (re)construct the route also covering the service points of the new request. 
As for preprocessing, in both the offline and the online scenarios, some common tasks are executed: 
	(a) The preprocessing phase of \ALGORITHM{CFLAT} was executed by computing optimal trees, so that the interpolation of the travel-times at destinations constitutes an $(1+\epsilon)$-approximation of the unknown time-dependent minimum-travel-time functions $\tau_h[o,d](t_o): L \times V \times T \mapsto \reals_{\geq 0}$, where $\epsilon=0.1$, $L$ is a subset of nodes (landmarks), $h=scooter$, and $T$ is a one-week period. In principle we could use the query algorithm \ALGORITHM{CFCA} to approximately compute time-dependent distances ``on the fly''. Nevertheless, since the graph size is small, we set $L=V$ so as to avoid executing \ALGORITHM{CFCA} and to improve the approximation guarantees of the provided travel-time values. 
	(b) Minimum distances $\lambda{[o,d]}: V \times V \mapsto \reals_{\geq 0}$ are computed with Dijkstra calls from all the nodes in $G_{PD}$, under the distance metric $\lambda$. 
	%Sanders' implementation of the sequence heap \cite{S19} was utilized for all the priority queues.

	%
We demonstrate some indicative results in Table~\ref{TABLE-1_TDINSERTION+TDPROPHET_DIST+L1-NORM}, which focuses on the distance metric and the $\ell_1$ objective.
The reported execution times of the online algorithms are average times per request. The table captures the resources spent: 
	total travel-time (h) and total-length (km) traveled by the workers to serve requests, 
	average (PathLen Avg) and variance (PathLen Var) of the workloads (measured in km) assigned to the workers. 
For the sake of a fair comparison, the human-curated subtours were translated into routes in such a way that all the interconnection paths are indeed distance-optimal paths, even if some of them are infeasible due to temporal constraint violations. This only works in favor of the baseline solutions. 
For the WB heuristic, we set $\theta =1.5$ and $\mu = 2$.
As shown in Table~\ref{TABLE-1_TDINSERTION+TDPROPHET_DIST+L1-NORM}, against the quality of the human-curated assignments, there is a clear improvement of all variants of \TDINSERTION{$^{heur}_{\lambda,\ell_1}$}, varying
	from $14.6\%$ up to $49.1\%$ decrease in total-length, and  
	from $13.5\%$ up to $48.9\%$ decrease in total travel-time.
The variants of \TDPROPHET{$^{heur}_{\lambda,\ell_1}$} provide an additional improvement over the corresponding variants \TDINSERTION{$^{heur}_{\lambda,\ell_1}$} by roughly $4\%$. The picture is similar also for the $\ell_2$ objective, as shown in
	Table~\ref{TABLE-2_TDINSERTION+TDPROPHET_DIST+L2-NORM} in the Appendix. 
	%Table 2 in the full version of the paper~\cite{2024-Kontogiannis-Paraskevopoulos-Zaroliagis-ARXIV}.
	%
Remarkably, $\ell_2$ does not necessarily provide better solutions, but it guarantees much less variance in the workloads, without the need of the WB heuristic.
	
As for the solutions of the relaxed MILP formulation of $\VRPPDSTCtd$, as shown in 
	Table~\ref{tab:offline-MILP-results} in the appendix, 
	%Table~5 in the full version of the paper, 
	%
	the involved solvers take hours to construct optimal solutions, even for small instances, whereas \TDINSERTION{} finds very good solutions within amortized time per request that is smaller by several orders of magnitude. E.g., for $12$ requests and $8$ workers, the branch-and-cut method of SCIP spending up to $6$ hours to find $31$ feasible solutions and, via them, in the next phase, the best time-dependent metric converted solution which is only $0.64\%$ better than the best of them in total-distance, within only $69ms$ per request.
	
\begin{table}[htb!]
\centerline{\includegraphics[width=0.6\paperwidth]{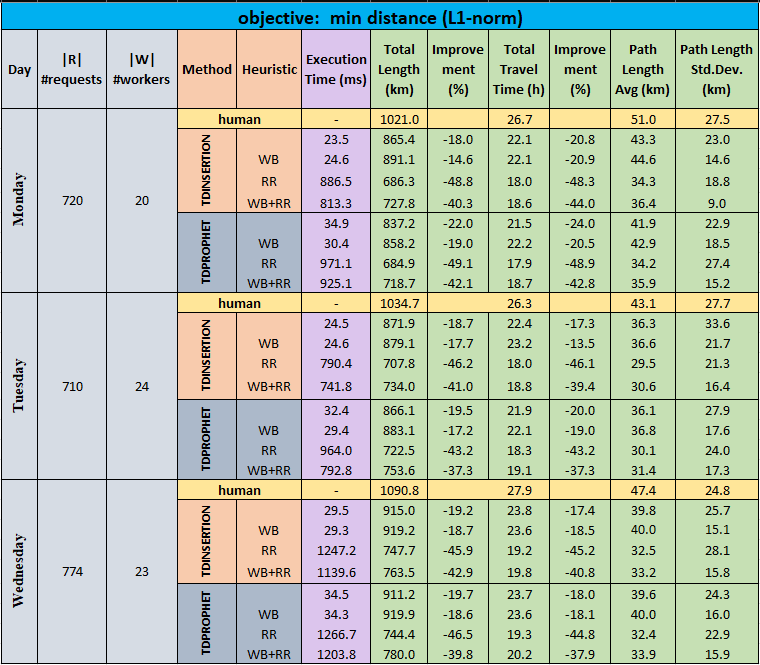}}
\caption{\label{TABLE-1_TDINSERTION+TDPROPHET_DIST+L1-NORM}%
		Experimentation of $\TDINSERTION{_{\lambda,\ell_1}}$ and $\TDPROPHET{_{\lambda,\ell_1}}$. 
}%CAPTION
\end{table}

\section{Concluding Remarks}

In this paper we introduced, implemented and engineered two insertion-based online schedulers for the \emph{time-dependent} variant $\VRPPDSTCtd$ of $\VRPPDSTC$, which were also experimentally evaluated on a real-world instance of food and shopping orders. In the future we plan to extend our online schedulers with more advanced local-search improvement heuristics, exploit them also by well known metaheuristics that are efficient for $\VRP{}$, such as the \ALGORITHM{ALNS} metaheuristic, and also to explore in more depth offline solvers which are custom-tailored to this particular problem. 

\bibliography{vrppd_references}

\appendix

%%%%%%%%%%%%%%%%%%%%%%%%%%%%%%%%%%%%%%%%%%%%%%%%%%%%%%%%%%%%%%%%%%%%%%%%%%%%%%%%%%
\section{Construction of a Relaxed MILP Formulation for $\VRPPDSTCtd$}
\label{sec:MILP-relaxation-of-VRPPDSTCtd}

Recall that each assignment of requests to workers for an instance of $\VRPPDSTCtd$ is represented as a collection $\left\{ S_w: w\in\mathcal{W}\right\}$ of vertex-disjoint (simple) paths in $G_{PD}=(\mathcal{V},\mathcal{E})$. Of course, in order to actually have a feasible solution in the end, we should translate each assignment $S_w$ into a (not necessarily simple) walk $\Pi_w$ to be followed by $w$ in the underlying road network $G=(V,E)$, by substituting each pair of consecutive points in $S_w$ (i.e., an arc in $G_{PD}$) with some cost-minimal interconnecting path of $G$. At this point, there are two options for the interconnection of the endpoints of each arc $e=uv\in\mathcal{E}$: Use in the road network $G$ either a travel-time-optimal (and distance-suboptimal) $(u,v)$-path, or a distance-optimal (and travel-time-suboptimal) $(u,v)$-path. 
When the cost-objective is based on the travel-time metric, all interconnecting paths for arcs of $\mathcal{E}$ are naturally minimum-travel-time paths in $G$, mainly due to the FIFO property of the metric. On the other hand, when the cost-objective is based on the distance metric, although the interconnecting paths for arcs of $\mathcal{E}$ should ideally be minimum-distance paths in $G$, such a choice might lead to infeasible solutions. We explain in subsection~\ref{sec:translating-assignments-to-routes-under-distance-metric} how we resolve this issue in such a way that, to the least, whenever there is a feasible solution of a given maximum number of serviced requests, one such solution should be found (even if it is suboptimal in the cost-objective).  

Before that, we first consider a simplified situation (cf.~\ref{sec:approximating-TD-travel-times}) where the travel-time metric consists of scalar values for the arcs in $G$ (i.e., it is time-independent). We then perceive all the temporal parameters (e.g., travel-times, arrival-times) as scalars, which can be precomputed. 
Abusing slightly the notation for the sake of simplicity, we use only the names of the temporal functions, without their explicit dependence on departure-time values from the tail of an arc or from the origin of a path, as the corresponding constants. For example, we write $\tau[\Pi_w]$ for the scalar approximation of the time-dependent path-travel-time function $\tau[\Pi_w](\workerStartTime_w)$. 
Given those (constant) travel-time values, in subsection~\ref{sec:MILP-Formulation-for-Relaxation-of-VRPPDSTCtd}) we provide a MILP formulation for the (time-independent) relaxation of the actual instance of $\VRPPDSTCtd$ that we wish to solve. This MILP is then fed to several MILP solvers for providing an offline solution (cf.~\ref{sec:MILP-solver}), to act as alternative baseline solutions for quantifying the quality of the provided heuristic solutions by our online solvers.
Of course, even the solutions provided by the offline solvers for the MILP relaxation are suboptimal solutions to the $\VRPPDSTCtd$ instance at hand. The challenge is exactly to adopt a time-independent travel-time metric which is somehow more informative than just considering the average, or the freef-flow arc-traversal times of the road segments and thus renders offline solutions closer to optimality.

%%%%%%%%%%%%%%%%%%%%%%%%%%%%%%%%%%%%%%%%%%%%%%%%%%%%%%%%%%%%%%%%%%%%%%%%%%%%%%%%%%
\subsection{Approximating Time-Dependent Travel-Times}
\label{sec:approximating-TD-travel-times}

This subsection concerns the determination of scalar travel-time values to all the arcs in the PD-graph $G_{PD}$, when the global objective to consider is the minimization of (sums, or sums-of-squares of) travel-time along the actual servicing paths of the workers. In this case, each arc of $G_{PD}$ actually represents a minimum-travel-time interconnecting path for its endpoints, in the underlying road network $G$. 

Rather than simply considering only average (or, free-flow) travel-time values per arc in $G$ and then conducting shortest-path computations between the endpoints of arcs in  $G_{PD}$, as is typically the case, we construct a more meaningful static travel-time metric for the arcs of $G_{PD}$, which tries to be as close as possible to the actual time-dependent travel-time metric in $G$, taking into account that specific connections may only appear at specific parts of the workers' subtours. In particular, we construct relaxated time-indepdenent travel-times for all the arcs in $G_{PD}$ in three consecutive phases,
	\begin{description}
		
	\item[Phase 1:]
		We compute the actual earliest arrival-times (and thus, also the minimum travel-times) from the starting location of each worker to every of the pickup point, under the time-dependent metric. In particular, fix an arbitrary arc 
	$e = (u=\workerStartPoint_w, v=\pickupPoint_r) \in \Vstart\times\Vpickup$ 
from some worker's shift-start point to some request's pickup point. The departure-time is definitely $\workerStartTime_w$. Therefore, using the query algorithm \ALGORITHM{CFCA} of the \ALGORITHM{CFLAT} oracle, we can determine a $(1+\epsilon)$-approximation of the minimum travel-time value 
$\tau[\Pi_{w}^{u:v}]({\workerStartTime_w})$. The eventual scalar travel-time approximation for $e$ (when considered as candidate for first arc in some subtour) is defined as 
	$\tau_{e,h_w} = \tau_{h_w}[\Pi_{w}^{u:v}](\workerStartTime_w)  + \pickupServiceTime_r$, 
i.e., we add to the actual travel-time value the service-time at the pickup point $\pickupPoint_r$.
	
	\item[Phase 2:]
		We consider all the arcs of $G_{PD}$ emanating from pickup points, towards other pickup points or delivery points (acting as candidates for second, or even later arcs within subtours). To compute scalar approximations of their travel-times, we make calls of \ALGORITHM{CFCA} from each pickup point $\pickupPoint_r$ and each vehicle-type $h\in H_r$, towards all destinations in $\Vpickup\cup\Vdelivery$. As departure-times from $\pickupPoint_r$ we consider the maximum of its earliest pickup-time $\requestEarliestPickupTime_r$ and its earliest arrival-time from any worker with the specific vehicle type. The resulting (time-dependent) minimum travel-time values at the destinations, plus the service times at the destinations, determine the scalar approximations $\tau_{e,h}$, for all the arcs $(\pickupPoint_r,v) \in \mathcal{E} \cap \Vpickup\times(\Vpickup\cup\Vdelivery)$ and vehicle types $h\in H_r$. 
	
	\item[Phase 3:]
		We consider all the arcs of $G_{PD}$ emanating from delivery points, towards other pickup points, delivery points, or work-shift ending points (as candidates for third, or even later arcs within subtours). To compute scalar approximations of their travel-times, we make again calls of \ALGORITHM{CFCA} from any delivery point $\deliveryPoint_r\in\Vdelivery$ towards all destinations in $v\in \Vpickup\cup\Vdelivery\cup\Vend$. 
		%\hl
		As departure-time from $\deliveryPoint_r$ we consider the earliest arrival-time, among all eligible vehicle types $h\in H_r$, from the corresponding pickup point $\pickupPoint_r$, as it was computed in the second phase, since it definitely has to precede that delivery point.
		%}%HL	
		The resulting minimum arrival-times computed by these calls plus the service times (if any) at the destinations, determine the scalar approximations $\tau_{e,h}$ of the travel-times that we consider, for all the arcs $(\deliveryPoint_r,v)\in \mathcal{E} \cap \Vdelivery\times(\Vpickup\cup\Vdelivery\cup\Vend)$
		%\hl{
		and eligible vehicle-type $h\in H_r$.
		%}%HL 

	\end{description}
%%%%%%%%%%%%%%%%%%%%%%%%%%%%%%%%%%%%%%%%%%%%%%%%%%%%%%%%%%%%%%%%%%%%%%%%%%%%%%%%%%
\subsection{MILP Formulation for Relaxation of $\VRPPDSTCtd$}
\label{sec:MILP-Formulation-for-Relaxation-of-VRPPDSTCtd}

With the above mentioned static travel-time metric at hand, we may proceed with the construction of the relaxed MILP formulation of $\VRPPDSTCtd$.
Recall that each arc $(\workerStartPoint_{w_1},\pickupPoint_{r_1}) \in \Vstart\times\Vpickup$ and each arc $(\deliveryPoint_{r_2},\workerEndPoint_{w_2}) \in \Vdelivery\times\Vend$ may be ``traversed'' by workers $w_1$ and $w_2$ if and only if $r_1$ and $r_2$ were assigned to them, respectively.
The rest of the arcs in $\mathcal{E}$ may be ``traversed'' by any worker whose vehicle is eligible for the serviced requests at its endpoint(s). 
Therefore, some binary decision variables are employed to indicate the traversal of arcs by workers and the assignment of requests to workers: For each arc 
	$e\in\mathcal{E} \cap ( (\Vpickup\cup\Vdelivery)\times(\Vpickup\cup\Vdelivery) )$, each request $r\in\mathcal{R}$, and each worker $w\in\mathcal{W}$,
	$x_{e,w}$ indicates whether $w$ traverses $e$, and 
	$x_{r,w}$ indicates whether $r$ is assigned to $w$. 

We proceed with the definition of some constants for earliest arrival-times at nodes in $\mathcal{V}$ and modifications in vehicle-loads, as some worker $w$ traverses an arc $e=uv\in S_w$ towards a service node $v\in\Vpickup\cup\Vdelivery$. Recall that any feasible solution is a collection of subtours  $\{S_w:w\in\mathcal{W}\}$ which correspond to vertex-disjoint paths in $G_{PD}$. In particular, for each edge $e=uv\in\mathcal{E}$, the constant $q_e$ represents the change in a vehicle's load, when traversing $e$:
	if $v = \pickupPoint_r\in\Vpickup$ then $q_e = q_r$;
	if $v=\deliveryPoint_r\in\Vdelivery$, then $q_e = -q_r$;
	otherwise, $q_e = 0$.
Moreover, each request $r\in\mathcal{R}$ comes with a (large positive) profit $\sigma_r$, to be considered only when $r$ is assigned to some worker for its service. 
Finally, for each node $v\in\mathcal{V}$, the continuous variable $a_v$ captures 
	either the arrival-time of the unique worker (if any) who serves the corresponding request $r$ along its assigned route, when $v\in\{\pickupPoint_r,\deliveryPoint_r \subseteq\Vpickup\cup\Vdelivery$;
	or a worker's shift starting time $\workerStartTime_w$, when $v=\workerStartPoint_w \in \Vstart$ is its shift-starting node;
	or, the eventual arrival-time at the end of the entire route $\Pi_{w}$ of a worker, when $v=\workerEndPoint \in \Vend$ is its shift-ending  node. 
Along each arc $e=uv\in\mathcal{E}$, the values of the variables $a_u$ and $a_v$ should be compliant with the required time for the moving worker (with a particular vehicle type) to traverse $e$. 

For the sake of simplicity, we make here the assumption that all workers possess the same vehicle type, as is the case in our real-world data set. The proposed relaxed mixed integer linear program (MILP) for $\VRPPDSTCtd$ is shown in Figure~\ref{fig:VRPPD-MILP}. It can be easily extended to also cover the case of more vehicle types for the workers.

\begin{figure}[tbh]
\begin{mini*}[3]
  {}
  {\sum_{e \in \mathcal{E}, w \in \mathcal{W}} c_{e,w} x_{e,w} - \sum_{r \in \mathcal{R}} \sigma_r \sum_{w \in \mathcal{W}} x_{r,w}}
  {}{}
  \addConstraint{ 0 \leq \sum_{v \in \Vpickup} x_{uv,\gamma(u)} \leq 1,}{\forall u \in \Vstart}																{\quad(1)} 
  \addConstraint{\sum_{uv \in \mathcal{E} : u = \pickupPoint_r} x_{uv,w} = x_{r,w} }{,~~ \forall r \in \mathcal{R}, w \in \mathcal{W}}					{\quad(2)}
  \addConstraint{\sum_{uv \in \mathcal{E} : v = \pickupPoint_r} x_{uv,w} = x_{r,w}}{,~~ \forall r \in \mathcal{R}, w \in \mathcal{W}}						{\quad(3)}
  \addConstraint{\sum_{uv \in \mathcal{E} : u = \deliveryPoint_r} x_{uv,w} = x_{r,w}}{,~~ \forall r \in \mathcal{R}, w \in \mathcal{W}}					{\quad(4)}
  \addConstraint{\sum_{uv \in \mathcal{E} : v = \deliveryPoint_r} x_{uv,w} = x_{r,w}}{,~~ \forall r \in \mathcal{R}, w \in \mathcal{W}}					{\quad(5)}
  \addConstraint{ 0 \leq \sum_{w \in \mathcal{W}} x_{r,w} \leq 1}{,~~ \forall r \in \mathcal{R}}															{\quad(6)}  
  \addConstraint{\vertexStartTime_v \leq a_v \leq \vertexEndTime_v}{,~~ \forall v \in \mathcal{V}}														{\quad(7)}
  \addConstraint{ a_{\pickupPoint_r} \leq  a_{\deliveryPoint_r}}{,~~ \forall r\in\mathcal{R}}																{\quad(8)}  
  \addConstraint{ |a_v - a_u -\sum_{w\in \mathcal{W}} (T_{max}+\tau_{uv,w})x_{uv,w}|  \leq T_{max}}{, \forall uv \in \mathcal{E}}						{\quad(9)}
  \addConstraint{|q_v - q_u -\sum_{w\in \mathcal{W}} (Q_{max}+q_{uv})x_{uv,w}| \leq Q_{max}}{, \forall uv \in \mathcal{E}}							{\quad(10)}     
  \addConstraint{0 \leq q_v \leq \sum_{r \in \mathcal{R}, w \in \mathcal{W}}  x_{r,w} \vehicleCapacity_w}{,~~ \forall v \in \mathcal{V_\mathcal{P}}}	{\quad(11)}
  \addConstraint{ x_{e,w}, x_{r,w} \in \{0,1\}}{,~~ \forall e:uv \in \mathcal{E}, w \in \mathcal{W}, r \in \mathcal{R}}										{\quad(12)}  
  \addConstraint{ q_v = 0, v \in \Vstart\cup\Vend;~~ 0 \leq q_v \leq Q_{max}, v \in \Vpickup\cup\Vdelivery}{}											{\quad(13)}
\end{mini*}
\caption{\label{fig:VRPPD-MILP}%
	The relaxed MILP formulation of $\VRPPDSTCtd$.
}%CAPTION
\end{figure}

The objective function seeks as a primary goal to maximize the number of served requests (recall the large positive values for the $\sigma_r$ parameters) and, as a secondary goal, to minimize the aggregate travel cost for having the selected requests served by the workers. Towards this direction, we construct the objective as the sum of two terms. 
The first term accounts for the aggregate cost to serve all the accepted requests, as determined by the sum of costs $c_{e,w}\cdot x_{e,w}$ for those arcs of $G_{PD}$ which are used in the solution. The value of the coefficient $c_{e,w}$ depends on the metric that is considered for the objective, i.e., it is associated with either the arc-length $\lambda_e$, or with the (approximate) arc-travel-time $\tau_{e,w}$ (but excluding the embedded service times).  
The second term determines the negative of the aggregate profit for serving requests: For each request $r\in \mathcal{R}$, the coefficient $\sigma_{r}$ denotes the ``profit'' for having $r$ served by some worker. These coefficients are set to a sufficiently large value (based on an upper bound to the worst possible path-travel-time or to the maximum length from any origin towards any destination), so as to enforce the service of as many requests as possible. The sum of profits for all the served requests is then subtracted from the overall service cost. 

As for the constraints of the MILP: 
	(1) ensures that any worker's subtour may start with a move towards at most one pickup point.
	(2-5) enforce that each worker may depart from / enter the pickup / delivery)  point of some request towards / from any other node, only if the corresponding request is assigned to her.
	(6) ensures that any request $r$ is served by at most one worker.
	(7) enforces the arrival time $ a_v$ at each node $v$ is in the allowable time window, where: 
			(a) for $v\in\{\workerStartPoint_w,\workerEndPoint_w\}$, ${\vertexStartTime_v}=\workerStartTime_w$ and $\workerEndTime_v = \workerEndTime_w$; and 
			(b) for $v\in\{\pickupPoint_r,\deliveryPoint_r\}$, $\vertexStartTime_v =\requestEarliestPickupTime_r$ and ${\vertexEndTime_v}=\requestLatestDeliveryTime_r$. 
	(8) ensures that the pickup-time point of $r$ precedes the delivery-time point of $r$. 
	(9) stipulates that if an arc $e=uv\mathcal{E}$ is traversed by some worker $w$, then the arrival time $a_v$ at $v$ must be the result of the arrival-time $a_u$ at $u$ plus $w$'s travel-time $\tau_{e,w}$ along $e$. 
	(10) stipulates that if an arc $e=uv\in\mathcal{E}$ is traversed by some worker $w$, then the change of $w$'s vehicle-load $q_v$ at $v$ results from the vehicle-load $q_u$ at $u$ plus the pickup-load / minus the delivery-load $q_{\rho(v)} = |q_e|$ that corresponds to the request $\rho(v)$. 
	(11) ensures that the vehicle's load at any pickup-node $v$ assigned to $w$, never exceeds the vehicle's maximum capacity $\vehicleCapacity_w$. 
The constants $T_{max}$ and $Q_{max}$ are the result of applying the \emph{big-M linearization} method, and their values, in direct dependence on the problem instance, are selected as the maximum distance $|a_v -a_u|$, $|q_v-q_u|$, $\forall uv\in\mathcal{E}$, e.g. a loose bound could be 
	$T_{\max} = \max_{w\in\mathcal{W}}(\workerEndTime_w - \workerStartTime_w)$ and 
	$Q_{\max} = \max_{w\in\mathcal{W}}\vehicleCapacity_w$. 
	Note that in (9) and (10) only the lower bounds are tight; the upper bounds are relaxed, $T_{max}$ and $Q_{max}$ theoretically can also be perceived as $\infty$. Their actual values are determined in relation to the rest of the constraints. This is done on purpose, especially for the arrival decision variables, because those variables in pickup-nodes have to be increased to reach the earliest pickup-times in the case of a non-zero buffer time.

%%%%%%%%%%%%%%%%%%%%%%%%%%%%%%%%%%%%%%%%%%%%%%%%%%%%%%%%%%%%%%%%%%%%%%%%%%%%%%%%%%
\subsection{Translating Assignments to Routes under the Distance Metric} 
\label{sec:translating-assignments-to-routes-under-distance-metric}

It is important to note at this point that for the MILP formulation provided in Figure~\ref{fig:VRPPD-MILP}, which was constructed on top of the PD graph with arc-costs equal to the minimum travel-times of their endpoints in the underlying road graph $G$, the optimal solutions indeed maximize the aggregate profit for serving the accepted requests and, at the same time, minimize the aggregate service cost for having the workers on the move, when the cost for each arc $uv\in \mathcal{E}$ is indeed measured by the total travel-time of the workers from $u$ to $v$ in the underlying road graph $G$. 

\begin{figure}[tbh]
	\centerline{\includegraphics[width=0.4\paperwidth]{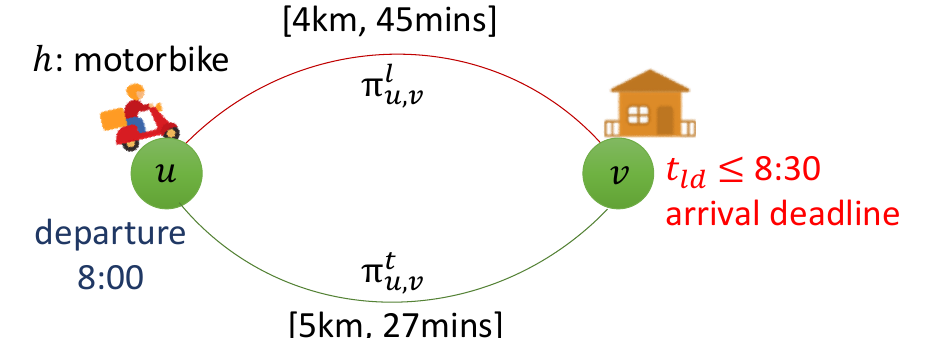}}
	\caption{\protect\label{fig:infsRisk}%
		Finding a feasible solution to minimize distance while respecting the time constraints.
		The worker arrives at $u$ at 8:00. 
		The arrival-time at node $v$ via the distance-optimal path $\pi_{u,v}^{l}$ is 8:45, i.e., too late w.r.t. the latest-delivery-time deadline (8:30). 
		On the other hand, the arrival-time at $v$ via the travel-time optimal path $\pi_{u,v}^{t}$ is 8:27, i.e., catching up the delivery deadline, at the cost of a slightly longer distance to travel.
	}%CAPTION
\end{figure}

Unfortunately, when the secondary objective (the aggregate service cost) is measured by the total distance traveled by the workers, it is no longer true that the consideration of a route of consecutive distance-optimal paths for implementing a given subtour is the right choice for implementing a worker's subtour. This then causes a crucial dilemma: \emph{which weights should be considered for the arcs of $\mathcal{E}$?} For example, in Figure~\ref{fig:infsRisk} the worker can move from $u$ to $v$ along one of two $uv$-routes in the underlying road network $G$, but the upper route is distance-optimal but time-infeasible and the lower route is travel-time optimal but distance-suboptimal. Then, for the unique arc $uv\in\mathcal{E}$, using the distance-optimal (but travel-time-suboptimal) weights would lead to a MILP formulation where the single request is impossible to serve, whereas using the travel-time-optimal (but distance-suboptimal) weights for the arcs in the PD graph would certainly provide a feasible solution, whose distance-related service cost  may be far from being optimal. 
Recall that our primary objective is to have a maximum profit by the accepted requests for service (e.g., to have as many requests served, as possible). Nevertheless, we wish (given that) to move towards optimizing also the distance-related service cost, even though we cannot possibly reach it. Towards this direction, we change the PD graph as follows:
For each $e = uv \in \mathcal{E}$, we attach two Pareto-optimal routes in the underlying road graph $G$, 
	a  distance-optimal (but travel-time-suboptimal) $uv$-route $\pi_{u,v}^{l}$, and 
	a travel-time-optimal (but distance-suboptimal) $uv$-route $\pi_{u,v}^{\tau}$, per type of vehicle. 
This way, our (updated for the distance-optimal service cost objective) MILP formulation tries to find either a distance-optimal and spatiotemporally feasible solution, or at least a distance-suboptimal feasible solution that employs the cheapest (w.r.t. extra distance to be traveled) subset of travel-time optimal connections so as to guarantee feasibility. Of course, this is still not the required distance-optimal solution for the maximum profit for the accepted requests, because it might be the case that some connecting paths in the road graph which are suboptimal for both distance and travel-time criteria might be preferable. Nevertheless, the primary goal of maximing the profit of accepted requests is now achieved.
 
%%%%%%%%%%%%%%%%%%%%%%%%%%%%%%%%%%%%%%%%%%%%%%%%%%%%%%%%%%%%%%%%%%%%%%%%%%%%%%%%%%
\subsection{Offline Solvers for Relaxed MILP Formulation of $\VRPPDSTCtd$}
\label{sec:MILP-solver}

The core method used for solving the MILP formulation of $\VRPPDSTC$ is based on built-in implementations in SCIP~\cite{BestuzhevaEtal2021,scip} and Gurobi \cite{gurobi} of the \emph{branch-and-cut} method \cite{BranchNCut}. 
Since the produced solution has taken into account, not the actual time-dependent travel-time metric, but a time-independent approximation metric for it, as was already explained in Section~\ref{sec:approximating-TD-travel-times}, we have to cross-check that the produced solution indeed respects all the spatiotemporal constraints of the instance. This is done as follows:
We examine up to 10 of the best feasible solutions found from the MILP solver. Each solution is a set of subtours $\left\{ S_w: w \in \mathcal{W}\right\}$. For each worker $w \in \mathcal{W}$ and the corresponding subtour $S_w$ which dictates the visiting order of the service points for all the requests assigned to her, we recompute the (now time-dependent) optimal interconnecting paths for consecutive points of $S_w$. The spatial constraints (relating to vehicle capacities) are certainly preserved. As for the temporal constraints, we recheck along the subtour if any arrival-time at a delivery service or shift-ending node now violates a temporal constraint, thus rendering the particular subtour infeasible. In such a case, we try to "repair" the route of $S_w$ in the following way: at each $v_x\in\Pi_w \cap \{\Vdelivery \cup \Vend\}$, where $\alpha(v_x) > \requestLatestDeliveryTime_{\rho(v_x)}$ or $\alpha(v_x) > \workerEndTime_w$ (i.e. a latest delivery or shift-ending deadline is violated), we traverse the route $\Pi_w^{v_1:v_x}$ backwards, i.e. from $v_x$ up to $v_1$, and successively any contained intermediate subpath in $\Pi_w$ that was computed for length-minimization is replaced by a corresponding optimal subpath that minimizes the travel-time. If there is no length-optimal path or the applied replacements eventually are not enough to deal with the deadline violations, then the whole solution is rejected, and an additional constraint to block the selection of the infeasible subtours in $G_{PD}$ is added to the MILP formulation (Figure~\ref{fig:VRPPD-MILP}). This process is applied to any MILP solver's examined feasible solution. At the end, if there is no time-dependent-metric converted feasible solution, then there is the possibility to use the new MILP formulation (with the added constraints) to solve the problem again for computing new solutions that will hopefully overcome the deadline violations.

%%%%%%%%%%%%%%%%%%%%%%%%%%%%%%%%%%%%%%%%%%%%%%%%%%%%%%%%%%%%%%%%%%%%%%%%%%%%%%%%%%
\section{Demand-Forecasting Procedure for \TDPROPHET{}}
\label{sec:TDPROPHET-demand-forecasting}

\begin{figure}[tbh]
	\centerline{\includegraphics[width=0.6\paperwidth]{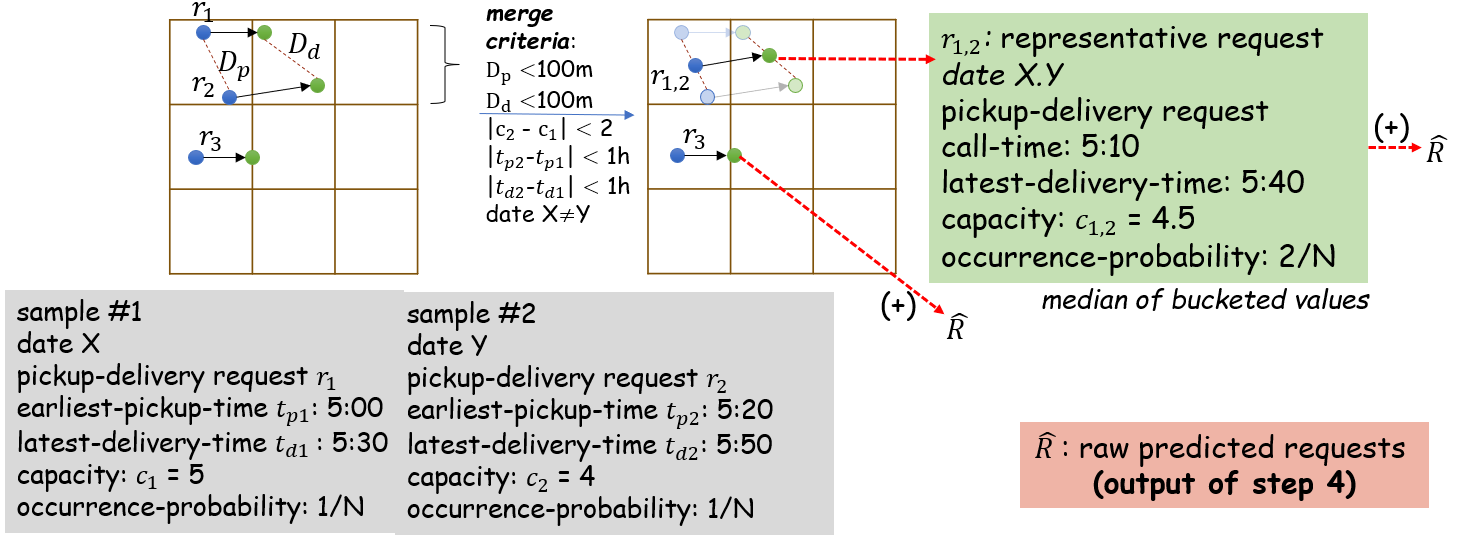}}
	\caption{\protect\label{fig:prediction_generation_process}%
Statistical processing of request data history. Example of merging two requests (on different days) with "similar" attributes to make predictions with a probability in relation to the number of the sampled days. 
}%CAPTION
\end{figure}

The major challenge is how to obtain accurate predictions, as this is critical for the efficiency of the \TDPROPHET{}.  Although the prediction process is beyond the scope of the present paper, we shortly describe a particular demand-prediction process, which was adopted in the framework of a research project in which our group participated~\cite{i-deliver}. It is based on some statistical preprocessing of historical data on real-world request sequences in the operational area of the data set that we consider, in order to provide meaningful demand forecasts. The preprocessing includes the following steps: 
	(i) A road network partition is set over $G$, creating $k$ cells, each one covering a $DxD$ $m^2$ area; 
	(ii) Each operational day is divided into 24 or more time windows, and each cell of the previous step is integrated into these time intervals, forming a spatiotemporal 3D grid partition of $G$; 
	(iii) From $N$ sample-days, the pickup-delivery requests made during them, are placed in the corresponding cell in the 3D grid; 
	(iv) An iterative merging process of merging and keeping the average or median is executed as follows: 
			for any pair of requests, 
				if they correspond to different days but their pickup and (time-location) events in the same cells, 
					and have a sufficient similarity in the rest attributes, 
				then they are merged (e.g., see Figure~\ref{fig:prediction_generation_process}). 
The stronger the merging process takes place, the better the demand trends are highlighted, providing generic request predictions between areas (rather than explicit locations and times) that have a high probability of occurring as recurrent events.

%%%%%%%%%%%%%%%%%%%%%%%%%%%%%%%%%%%%%%%%%%%%%%%%%%%%%%%%%%%%%%%%%%%%%%%%%%%%%%%%%%
\section{Detailed Presentation of Experimental Results}
\label{sec:detailed-experiments-in-appendix}

In this section we present more statistics for the experimental evaluation of our online algorithms. The main results are shown in Tables
	\ref{TABLE-1_TDINSERTION+TDPROPHET_DIST+L1-NORM},
	\ref{TABLE-2_TDINSERTION+TDPROPHET_DIST+L2-NORM},
	\ref{TABLE-3_TDINSERTION+TDPROPHET_TRAVEL-TIME+L1-NORM}, and
	\ref{TABLE-4_TDINSERTION+TDPROPHET_TRAVEL-TIME+L2-NORM}. 
The table captures the resources spent: 
	total travel-time (h) and total-length (km) traveled by the workers to serve requests, 
	average (PathLen Avg) and variance (PathLen Var) of the workloads (measured in length) assigned to the workers. 
It is noted that, for the sake of fair comparison (since the dataset's real-time raw worker routes might have had additional delays), the worker routes based on manual assignments were edited so that the interconnection paths between consecutive work-shift start points, service points and work-shift end points, are indeed distance-optimal routes, without changing the original subtours determined by the human operators. 
For the WB heuristic, the chosen parameters are $\theta =1.5$ and $\nu = 2$.

Considering the online scenario for $\VRPPDSTCtd$, two main reasons arise for getting a worse cost: 
	(a) the suboptimal assignments and the result of them; 
	(b) the detours that need to be entered on they fly, as new requests are revealed.

The exact solution of the MILP formulation of the static $\VRPPDSTC$ instance has also been examined as a feasible solution of $\VRPPDSTCtd$, see Table~\ref{tab:offline-MILP-results}. The construction of an exact solution for the (static) $\VRPPDSTC$ instance by the branch-and-cut method, even on small problem sizes, requires several hours of execution time. E.g., for a subset of 10 requests and at most 5 workers, the \TDINSERTION{} algorithm finds a satisfactory solution, with at most $10\%$ worse cost (both for the distance and for the travel-time objectives) within $3ms$. On the other hand, branch-and-cut method requires $1$min, which already indicates the huge gap in the execution-times of the two methods.

\begin{table}[tbh]
\centerline{\includegraphics[width=0.6\paperwidth]{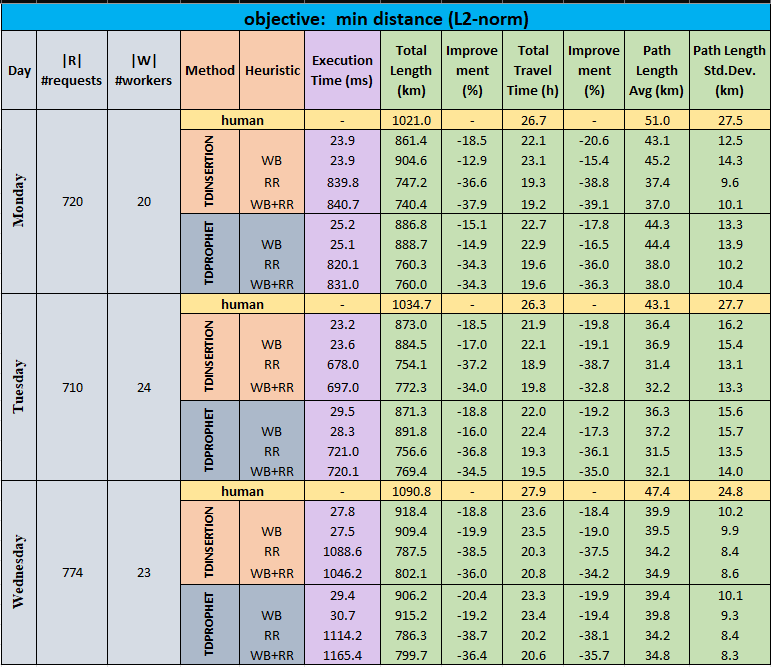}}
\caption{\protect\label{TABLE-2_TDINSERTION+TDPROPHET_DIST+L2-NORM}%
		Experimentation of variants of \TDINSERTION{$^{heur}_{\lambda,\ell_2}$} and \TDPROPHET{$^{heur}_{\lambda,\ell_2}$} (i.e., distance metric and $\ell_2$ global objective).  
}%CAPTION
\end{table}

\begin{table}[tbh]
\centerline{\includegraphics[width=0.6\paperwidth]{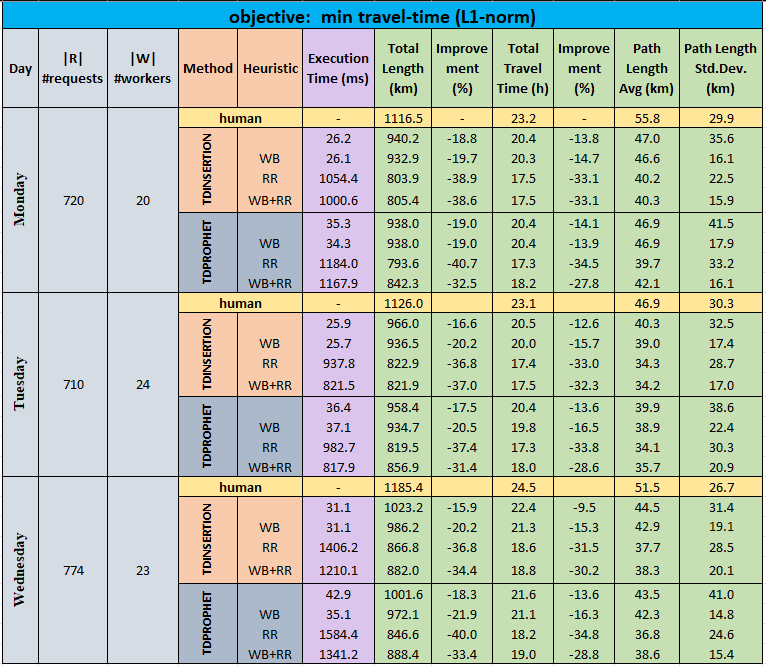}}
\caption{\protect\label{TABLE-3_TDINSERTION+TDPROPHET_TRAVEL-TIME+L1-NORM}%
		Experimentation of variants of \TDINSERTION{$^{heur}_{\tau,\ell_1}$} and \TDPROPHET{$^{heur}_{\tau,\ell_1}$} (i.e., travel-time metric and $\ell_1$ global objective).  
}%CAPTION
\end{table}

\begin{table}[tbh]
\centerline{\includegraphics[width=0.6\paperwidth]{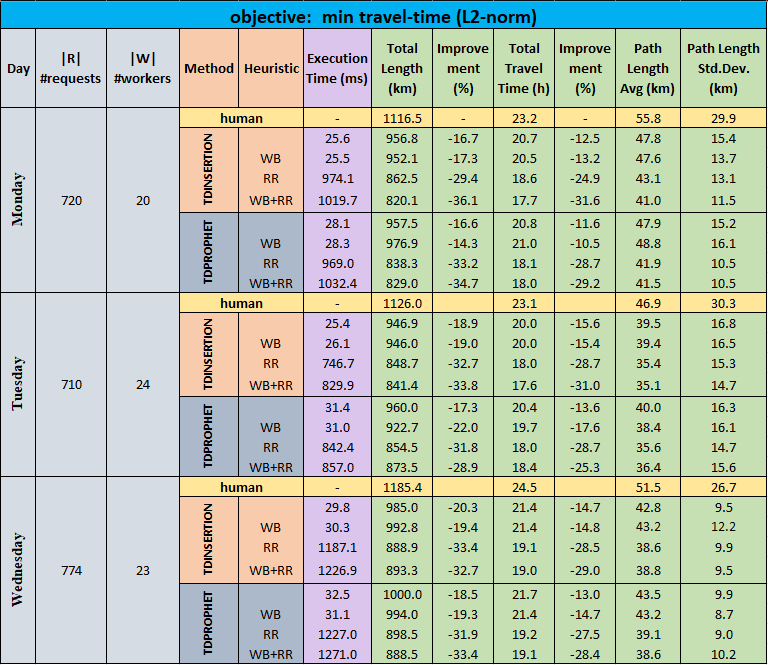}}
\caption{\label{TABLE-4_TDINSERTION+TDPROPHET_TRAVEL-TIME+L2-NORM}%
		Experimentation of variants of \TDINSERTION{$^{heur}_{\tau,\ell_2}$} and \TDPROPHET{$^{heur}_{\tau,\ell_2}$} (i.e., travel-time metric and $\ell_2$ global objective).  
}%CAPTION
\end{table}

Tables
	\ref{TABLE-2_TDINSERTION+TDPROPHET_DIST+L2-NORM}, 
	\ref{TABLE-3_TDINSERTION+TDPROPHET_TRAVEL-TIME+L1-NORM} and
	\ref{TABLE-4_TDINSERTION+TDPROPHET_TRAVEL-TIME+L2-NORM} 
demonstrate the performance of our online schedulers $\TDINSERTION{_{\kappa,\nu}}$, for 
$(\kappa,\nu) \in\{ (\lambda,\ell_2), (\tau,\ell_1), (\tau,\ell_2) \}$.
\begin{figure}[tbh]
\centerline{\includegraphics[width=0.5\paperwidth]{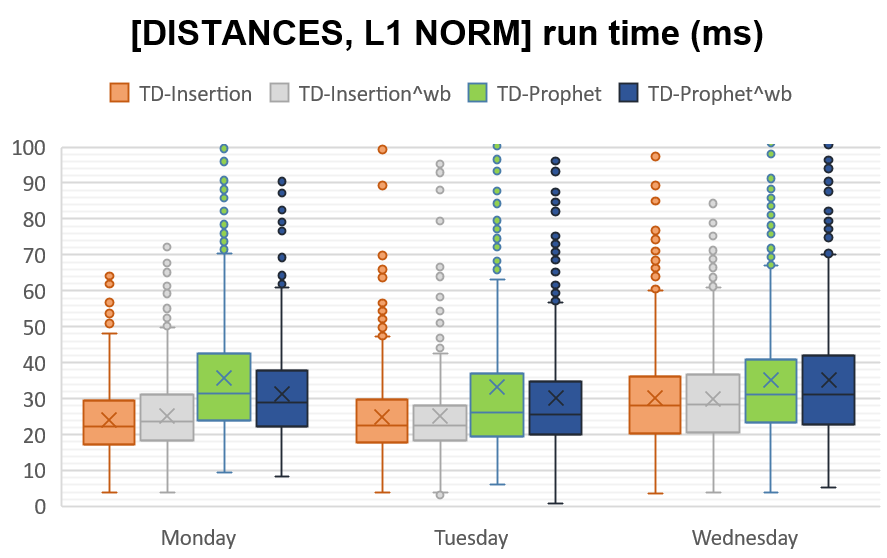}}
\caption{\label{fig:run_times_TDINSERTION+TDPROPHET_variants_1}%
		Statistics of execution times (ms) for the online schedulers 
		\(
			\TDINSERTION{_{\lambda,\ell_1}}, 
			\TDINSERTION{^{wb}_{\lambda,\ell_1}}, 
			\TDPROPHET{_{\lambda,\ell_1}},
			\TDPROPHET{^{wb}_{\lambda,\ell_1}}
		\)	 
		on $\VRPPDSTCtd$  instances. 
}%CAPTION
\end{figure}

\begin{figure}[tbh]
\centerline{\includegraphics[width=0.5\paperwidth]{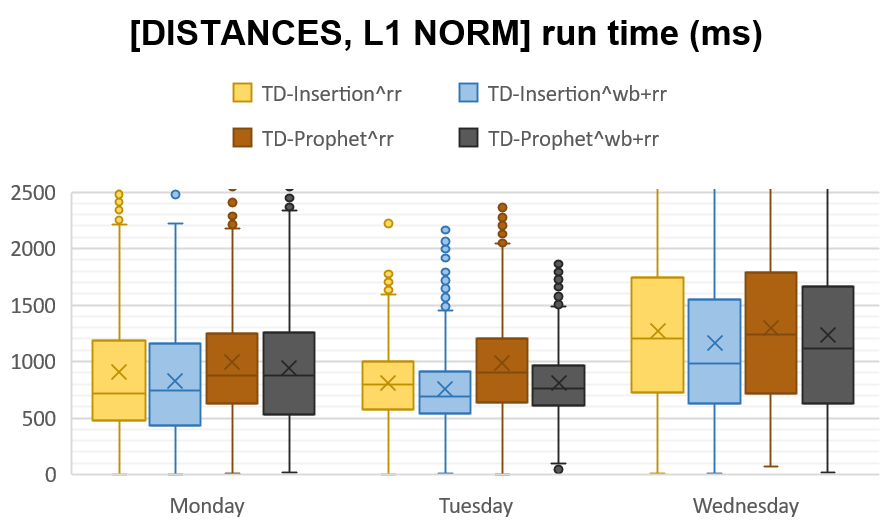}}
\caption{\protect\label{fig:run_times_TDINSERTION+TDPROPHET_variants_2}%
		Statistics of execution times (ms) of the online schedulers 
		\(
			\TDINSERTION{^{rr}_{\lambda,\ell_1}}, 
			\TDINSERTION{^{wb+rr}_{\lambda,\ell_1}}, 
			\TDPROPHET{^{rr}_{\lambda,\ell_1}},
			\TDPROPHET{^{wb+rr}_{\lambda,\ell_1}}
		\)	 
		for $\VRPPDSTCtd$  instances. 
}%CAPTION
\end{figure}
Figures~\ref{fig:run_times_TDINSERTION+TDPROPHET_variants_1} and \ref{fig:run_times_TDINSERTION+TDPROPHET_variants_2} provide the statistics of execution-times per request, for all the variants of \TDINSERTION{} and \TDPROPHET{}. 
\begin{figure}[tbh]
\centerline{\includegraphics[width=0.75\paperwidth]{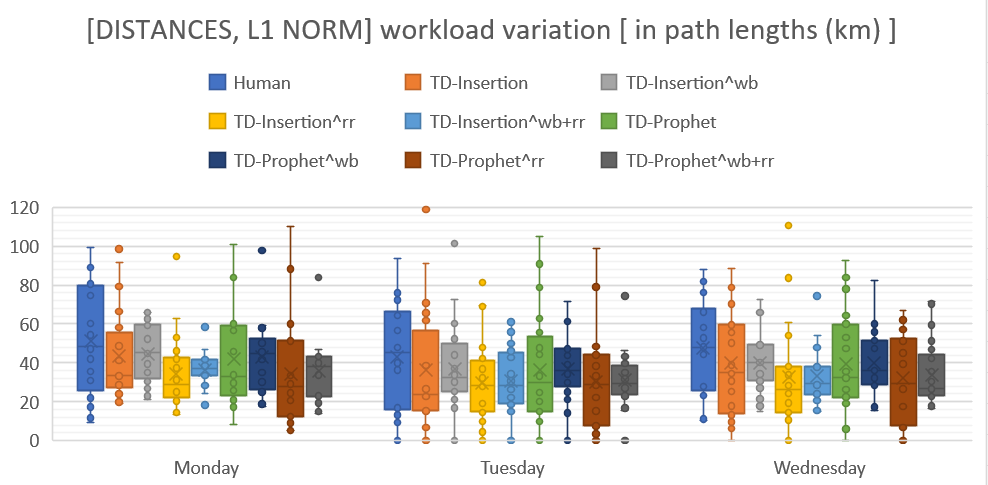}}
\caption{\protect\label{fig:workload-distributions_DIST+L1}%
		Workload distribution (length per worker, in km) of $\TDINSERTION{_{lambda,\ell_1}}$. 
}%CAPTION
\end{figure}
\begin{figure}[tbh]
\centerline{\includegraphics[width=0.75\paperwidth]{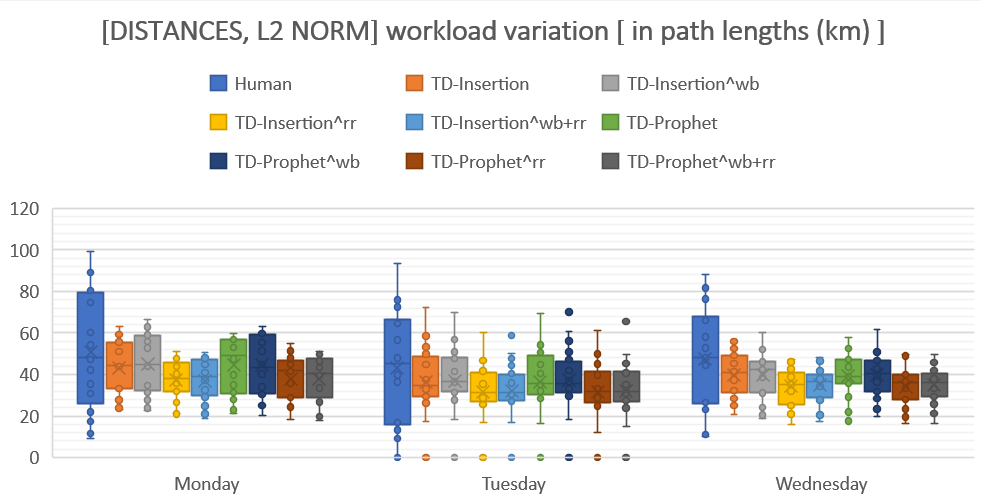}}
\caption{\protect\label{fig:workload-distributions_DIST+L2}%
		Workload distribution (length per worker, in km) of $\TDINSERTION{_{\lambda,\ell_2}}$. 
}%CAPTION
\end{figure}
Figures~\ref{fig:workload-distributions_DIST+L1} and \ref{fig:workload-distributions_DIST+L2} demonstrate the statistics of workloads of the solutions provided by all the variants of $\TDINSERTION{^{heur}_{\kappa,\ell_1}}$ and $\TDINSERTION{^{heur}_{\kappa,\ell_2}}$, respectively.
	
\begin{table}[tbh]
\centerline{\includegraphics[width=0.5\paperwidth]{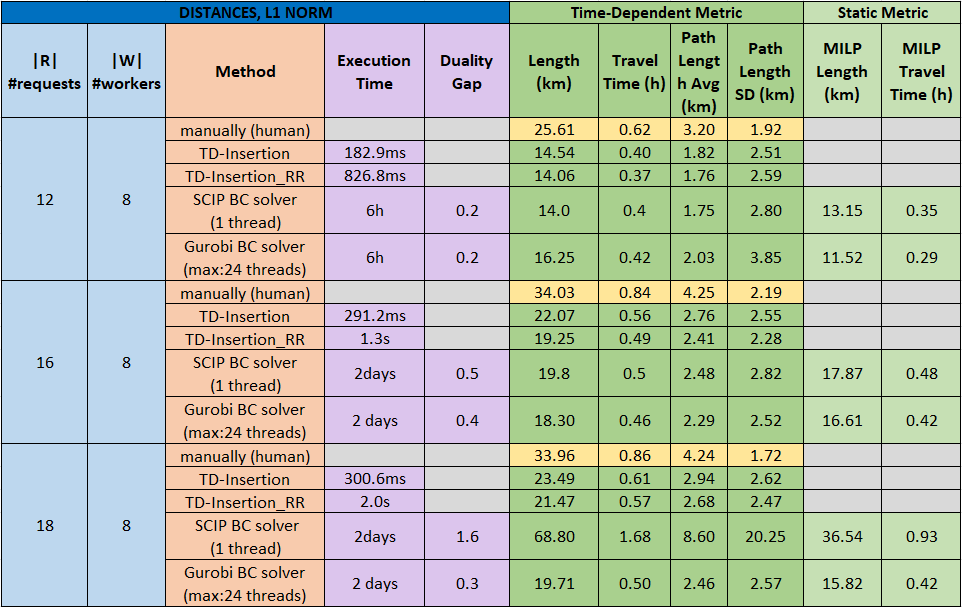}}
\caption{\protect\label{tab:offline-MILP-results}%
	Exact problem solving of mixed integer linear programs corresponding to pickup-delivery problem snapshots. 
}%CAPTION
\end{table}
Finally, Table~\ref{tab:offline-MILP-results} compares the efficiency (execution time) and effectiveness (solution quality) of the feasible solutions constructed using the Branch-and-Cut solver of SCIP (with an upper bound of $3$h on the execution time) for the MILP formulation of the relaxed problem (with static travel-times) $\VRPPDSTC$. The comparison is with the human-curated solutions and the solutions provided by the $\TDINSERTION{_{\tau,\ell_1}}$ and the $\TDINSERTION{^{rr}_{\tau,\ell_1}}$ online schedulers.  For the SCIP solver, we consider the best ten feasible solutions to  $\VRPPDSTC$, and then only keep the best of them under the time-dependent travel-time metric. 

\end{document}